\documentclass[twocolumn,twocolappendix]{aastex631}

\usepackage{graphicx,hyperref}	
\usepackage{amsmath,makecell}	
\newcommand{\cmmc}{\textsc{\small 21CMMC}}
\newcommand{\cmfst}{\textsc{\small 21cmFAST}}
\newcommand{\ttan}{\textsc{\small THESAN}}

\submitjournal{ApJ}

\shorttitle{21~cm few-shot GAN}
\shortauthors{Diao \& Mao}

\begin{document}

\title{Multi-fidelity emulator for large-scale 21~cm lightcone images: a few-shot transfer learning approach with generative adversarial network}

\author[0000-0001-7301-2318]{Kangning Diao}
\affiliation{Department of Astronomy, Tsinghua University, Beijing 100084, China}
\affiliation{Berkeley Center for Cosmological Physics, University of California, Berkeley, CA 94720, United States}

\author[0000-0002-1301-3893]{Yi Mao}
\affiliation{Department of Astronomy, Tsinghua University, Beijing 100084, China}

\correspondingauthor{Kangning Diao, Yi Mao}
\email{dkn20@mails.tsinghua.edu.cn (KD), ymao@tsinghua.edu.cn (YM)}



\begin{abstract}
Emulators using machine learning techniques have emerged to efficiently generate mock data matching the large survey volume for upcoming experiments, as an alternative approach to large-scale numerical simulations. However, high-fidelity emulators have become computationally expensive as the simulation volume grows to hundreds of megaparsecs. 
Here, we present a {\it multi-fidelity} emulation of large-scale 21~cm lightcone images from the epoch of reionization, which is realized by applying the {\it few-shot transfer learning} to training generative adversarial networks (GAN) from small-scale to large-scale simulations. Specifically, a GAN emulator is first trained with a huge number of small-scale simulations, and then transfer-learned with only a limited number of large-scale simulations, to emulate large-scale 21~cm lightcone images. We test the precision of our transfer-learned GAN emulator in terms of representative statistics including global 21~cm brightness temperature history, 2D power spectrum, and scattering transform coefficients. We demonstrate that the lightcone images generated by the transfer-learned GAN emulator can reach the percentage level precision in most cases on small scales, and the error on large scales only increases mildly to the level of a few tens of per cent. Nevertheless, our multi-fidelity emulation technique saves a significant portion of computational resources that are mostly consumed for generating training samples for GAN. On estimate, the computational resource by training GAN completely with large-scale simulations would be one to two orders of magnitude larger than using our multi-fidelity technique. This implies that our technique allows for emulating high-fidelity, traditionally computationally prohibitive, images in an economic manner.

\end{abstract}



\section{Introduction}

The epoch of reionization \citep[EoR; see, e.g.][]{Morales2010,Pritchard_2012} is a critical period in the history of our universe, marking the last phase transition. Despite its importance, EoR remains mysterious due to insufficient observations. A widely accepted picture of EoR is the bubble model \citep[e.g.][]{Furlanetto_2016}, where ionizing sources emit UV and X-ray photons, ionizing the surrounding intergalactic medium (IGM) and creating ionized bubbles. These bubbles then expand and merge, eventually occupying the entire universe by the end of EoR \citep{Chen_2019}. 

Several observations have been used to probe EoR, including optical depth measurement of the cosmic microwave background (CMB; e.g.\ \citealt{Planck_2020}), galaxy survey \citep[e.g.][]{Labbe_2022,Naidu_2022}, Ly$\alpha$ forest (e.g. \citealp{Morales}; {\citealp{2021ApJ...923..223Z}; \citealp{2023MNRAS.523.1399D}}), and 21~cm line \citep[e.g.][]{Furlanetto2006}. 
The 21~cm line due to the hyperfine spin-flip transition of atomic hydrogen is a particularly promising tracer, since it can directly probe the state of the IGM during reionization. Many radio telescopes are ongoing or under construction to measure the global 21~cm signal, e.g.\ EDGES \citep{Bowman_2018}, SARAS \citep{T_2021, Saras3}, or measure the spatial fluctuations of the 21~cm signal from the EoR, e.g.\ LOFAR \citep{van_Haarlem_2013}, MWA \citep{MWA_2013}, PAPER \citep{Parsons_2014}, HERA \citep{DeBoer_2017}, and SKA \citep{Koopmans_2015}. Moreover, the SKA has the potential for making the images of the IGM through the 21~cm emission directly.

In preparation for the new era with the 21~cm imaging, many techniques have been developed to extract information from observations. Specifically, the Monte Carlo Markov chain (MCMC) method, e.g.\ \cmmc\ code \citep{Greig_2017}, and likelihood-free inference (LFI; \citealt{Alsing_2019,Zhao_2022}), e.g.\ {\textsc{\small 21cmDELFI-PS}} \citep{Zhao_2022B} and {\textsc{\small Scatter-Net}} \citep{2024ApJ...973...41Z}, have been developed to infer reionization and astrophysical parameters from the 21~cm EoR signals. These methods require performing a large number of simulations, either for computing MCMC chains or for preparing training samples. These simulations range from the semi-numerical simulations, e.g.\  \cmfst\ \citep{Mesinger_2010,Murray_2020}, to radiative transfer simulations, e.g.\ \ttan \citep{Kannan_2021} and {\textsc{\small C$^2$-Ray}} simulations \citep{C2RAY,pyC2RAY}, with different levels of accuracy and computational costs. Given the large field of view of the next-generation telescopes, such as SKA and HERA, large-scale simulations are required to fully exploit the information from observations. However, all of these large-scale simulations are more or less computationally expensive, if not prohibitive. This bottleneck problem has inspired the development of emulators as an alternative approach to simulations.

Building emulators typically requires numerous training samples, which contradicts the original purpose of reducing computational costs. One possible solution is to build a data reservoir that gathers as many state-of-the-art simulations as possible. The publicly available CAMELS project \citep{Nararro_2022} and the LoReLi database \citep{2024A&A...683A..24M} are such successes that have demonstrated their impacts on emulator building. However, this issue turns out to be particularly serious as the simulation volume grows to be larger than hundreds of megaparsecs on each side, in that high-fidelity emulators have become computationally expensive in this case. To address this issue, the concept of multi-fidelity emulation \citep{Kennedy_2000,Ho_2021} has been proposed and guided the design of dataset \citep{2025arXiv250106296Y}.  In this approach, a large number of low-fidelity simulations -- i.e., low-cost with lower resolution or simpler algorithms -- are first used to train an emulator. The emulator is then calibrated with a small number of high-fidelity simulations, i.e.\ high-cost with higher resolution or more complicated algorithms. In this manner, the computational cost can be significantly reduced while still maintaining a reasonable output quality.

Machine learning (ML) has become a popular tool for astronomy in recent years, with diverse applications ranging from classifying models \citep[e.g.][]{Hassan_2017,Hassan_2018}, parameter inference \citep[e.g.][]{Shimabukuro_2017,Gillet_2019,Hassan_2020,Zhao_2022}, segmenting components (e.g.\ Sui et al. in prep) and generating images \citep[e.g.][]{Hassan_2022}. The Generative adversarial network (GAN; \citealt{Goodfellow_2014}) has emerged as a powerful ML model for generating quality images thanks to its fast generation speed and high image quality. {Comparing with deterministic emulators that are based on functions fitting with multi-layer perceptron \citep[e.g.][]{2024MNRAS.527.9977S,2024arXiv240703523C,2024MNRAS.527.9833B} or symbolic regression \citep[e.g.][]{2024arXiv240513680M,2024arXiv241014623S}, generative models such as GAN are effective in high-dimensional applications such as the cosmological fields, thus preserving high‑order and non‑Gaussian statistics. Meanwhile, generative models can capture uncertainties, which is suitable for cosmological fields with initial random conditions.} 
The GAN has been applied to the emulation of astrophysical images \citep[e.g.][]{2022JCAP...12..013Y,2019MNRAS.487L..24T,2019MNRAS.490.3134L,2021MNRAS.506..357Y} {and enhancing the simulation resolution \citep[e.g.][]{2021PNAS..11822038L,2021MNRAS.507.1021N,2024MNRAS.528..281Z,2025OJAp....8E..13Z,2023ApJ...958...21J}}. \citet{List_2020} demonstrated that the GAN, together with the approximate Bayesian computation (ABC) method, can be used to accurately estimate the reionization parameters. Furthermore, \citet{Sambatra_2022} showed an emulation of multi-field images with GAN, which preserves the cross-correlations between different fields. 

In this paper, we present a {\it multi-fidelity} emulation of large-scale 21~cm lightcone images from the EoR with GAN, as a trade-off between computational cost and emulation quality. Technically, this is achieved by applying the {\it few-shot transfer learning} method \citep[e.g.][]{Ojha_2021}, which allows for training a faithful GAN emulator with a limited number of samples and serves as the calibrating procedure in multi-fidelity emulation. Specifically, a GAN emulator is first trained with a large number of small-scale simulations, and then transfer-learned with only a limited number of large-scale simulations, to emulate large-scale EoR lightcone images. Our transfer-learned GAN emulator will be tested for precision in terms of several EoR statistics. As such, our GAN version of multi-fidelity emulation serves as a promising approach to generate data sets with high image quality and low computational cost.

When the manuscript of this paper was in preparation, diffusion models \citep{jonathanhodiffusion,song2021}, along with flow matching models with more general forward paths \citep{flowmatching}, appeared recently as new models that also have the ability to generate high-quality data without adversarial training. \citet{2023MNRAS.526.1699Z} applied the diffusion model to generate the images of 21~cm brightness temperature mapping as a case study to conduct a quantitative comparison between the denoising diffusion probabilistic model (DDPM) and StyleGAN2. While these state-of-the-art models are promising alternatives to GANs in generating accurate images, our work will present a successful example of multi-fidelity emulation of astrophysical images with GAN and therefore sheds light on such similar possible applications with other emulation techniques (e.g.\ the diffusion model and the flow matching model). 



The remainder of this paper is organized as follows. In Section \ref{sec:GAN}, we introduce our method for training a GAN emulator with limited data. In Section \ref{sec:data}, we summarize the astrophysical model for generating the data sets. In Section \ref{sec:ssgan}, we evaluate our small-scale GAN model with several statistics. We assess the precision of our final objective, the large-scale GAN, in Section \ref{sec:lsgan}, and make concluding remarks in Section \ref{sec:con}. We leave some technical details to Appendix~\ref{app:GAN} (on GAN architecture and configurations), {and Appendix~\ref{app:80} (on the result of training large-scale GAN only with 80 simulations without the multi-fidelity emulation technique).} Some of our results were previously summarized by us in a conference paper \citep{2023mla..confE..12D}. 

\section{GAN training via few-shot transfer learning}
\label{sec:GAN}

GAN is a type of generative model that is used to create new images based on a given data set. Among all types of generative models, GAN has the advantage of fast generation {comparing to diffusion models} and high image quality {comparing to normalizing flows}, which makes it suitable for emulators. {Moreover, GAN has the flexibility of choosing the loss function, allowing for injecting more inductive bias by altering the loss function design.} However, GAN may suffer from the so-called model collapsing problem, which means that the generated images lack diversity, especially when the data set size is limited. This means that GAN usually requires a large data set for training, which can be time-consuming and costly. In this work, we apply the idea of GAN few-shot transfer learning, aiming to train a GAN with as few training samples as possible, to reduce the computing resource requirement.

Our approach is a two-step process. First, we train our GAN with 120,000 small-scale images. The number of small-scale images is typically sufficient for GAN training, making our small-scale GAN immune to model collapse. We modify specific layers of small-scale GAN, making our GAN capable of generating large-scale images, i.e.\ creating a large-scale GAN. In the second step, we train our large-scale GAN with 320 large-scale images, with a patchy-level generator, a layer-frozen \citep[FreezeD;][]{Mo_2020} multi-scale discriminator, and the cross-domain correspondence \citep[CDC;][]{Ojha_2021} to maintain the model diversity. The details of our approach are given below. 

\subsection{StyleGAN2}
\label{sec:stygan}
\begin{figure*}
 \includegraphics[width=\linewidth]{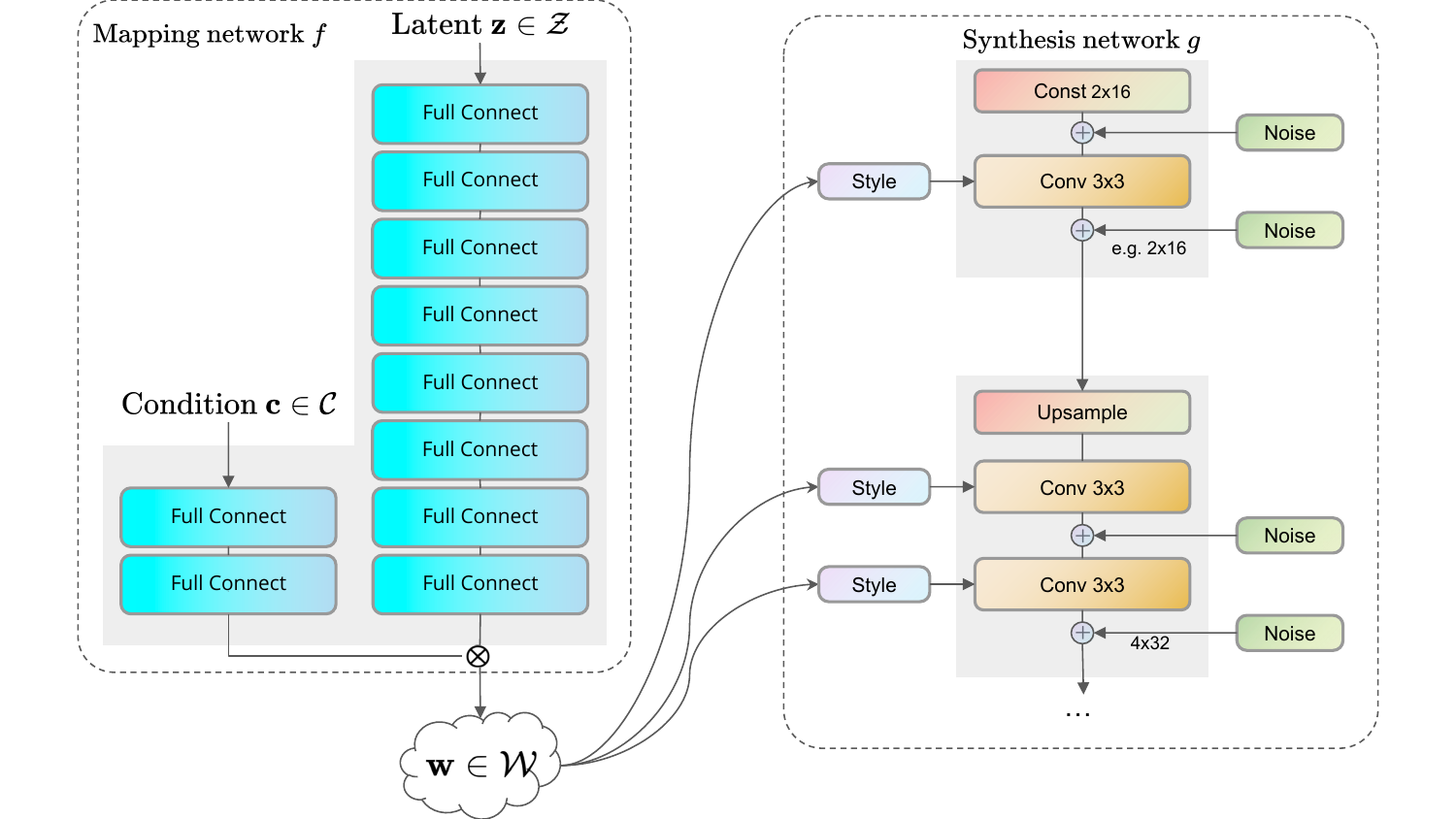}
 \caption{An illustration of the StyleGAN2 generator architecture. Our generator consists of a mapping network $f$, which modifies the convolution kernel according to astrophysical parameters and random vectors, and a synthesis network $g$, which generate images progressively, with noise injection for multiple times.}
 \label{fig:stylegan}
\end{figure*}
A GAN typically consists of two parts, a generator $G$, and a discriminator $D$, both of which are deep neural networks. The most naive form of the conditional GAN loss function is the so-called adversarial loss, 
\begin{equation}
    \mathcal{L}_{\rm adv} = {\log(1{- D(G(\mathbf{z},\mathbf{c})|\mathbf{c}))}) + \log({ D(\mathbf{x}|\mathbf{c})})}\,,
\end{equation}
Here the generator $G$ is a function that outputs an emulated image with the input of a random vector $\mathbf{z}$ and a set of astrophysical parameters $\mathbf{c}$. The discriminator $D$, given an image and the corresponding astrophysical parameters $\mathbf{c}$ as input, makes a decision on whether the input image is real or not and empirically outputs the value of zero for the fake and unity for the real. $\mathbf{c}$ is the condition, e.g.\ the astrophysical parameters in our case. $\mathbf{z}$ is a random vector that provides stochastic features. $\mathbf{x}$ is the real image sample in our training set. 
The training objective is finding the optimal $G$ and $D$ models, as labeled by $(G^*,D^*)$, obtained by
\begin{equation}
    (G^*,D^*) = \arg \min\limits_{G} \max \limits_{D} \mathbb{E}_{\mathbf{z}\sim p(\mathbf{z}),\mathbf{x}\sim p(\mathbf{x})} \mathcal{L}_{\rm adv}\,.
\end{equation}
Here $p(\mathbf{z})$ is the probability distribution of $\mathbf{z}$, modeled as a multivariate diagonal Gaussian distribution, and the distribution of real images $p(\mathbf{x})$ {is approximated by the empirical distribution of our training set}. $\mathbb{E}$ means taking expectations over distributions. In practice, samples from $p(\mathbf{x})$ and $p(\mathbf{z})$ are used to obtain an empirical estimation of the expectation with maximum steps for $D$ and minimum steps for $G$  in turn (denoted by $\arg \min\limits_{G} \max \limits_{D}$). 

In this work, we employ \texttt{StyleGAN2} \citep{Karras_2019}, the second version of the state-of-the-art GAN model, as the GAN architecture. We illustrate the generator architecture in Figure~\ref{fig:stylegan}. The discriminator is the commonly used ResNet \citep{He_2015} architecture. Our generator consists of two parts. First, a mapping network $f$ takes the set of astrophysical parameters $\mathbf{c}$ and a random vector $\mathbf{z}$ and returns a style vector $\mathbf{w}$. Secondly, a synthesis network $g$ uses the style vector $\mathbf{w}$ to shift the weights in the convolution kernels, and Gaussian random noise is injected into the feature map right after each convolution to provide variations in the detail of the emulated map. The main structure of $g$ keeps the form of progressively growing GAN, which generates the map with a poor resolution, e.g.\ $2\times8$, and upsamples after convolutions until reaching the desired size. Our realization is publicly available in this GitHub repo\footnote{\url{https://github.com/dkn16/stylegan2-pytorch}, which is based on \url{https://github.com/rosinality/stylegan2-pytorch}.}. This architecture is interesting because it is similar to {the 21cmFAST model}: {while the 21cmFAST model evolves the Gaussian initial condition with Lagrangian perturbation and excursion set to obtain the 21~cm brightness temperature field, the GAN model convolves the Gaussian random noise with convolutional kernels to output the same field. While the reionization parameters affect the postprocessing of the initial condition, the same set of parameters only modify the convolutional kernel in the GAN model, rather than acting directly on the random field.}

Beyond the simple form of the loss function, regularization is put on the generator and discriminator, respectively. An $r_1$ loss $\mathcal{L}_{r_1}$ \citep{Lars_2018} is applied to the discriminator to improve the sparsity of the weight matrices, alleviating overfitting. A path-length loss is applied to the generator, which has the form of
\begin{equation}
    \mathcal{L_{\mathrm{path}}} = \left[\Big|\Big|\frac{\partial {g}(\mathbf{w})}{\partial \mathbf{w}}^{T}{g}(\mathbf{w})\Big|\Big|_2-a\right]^2\,.
\end{equation}
{where $g$ is the synthetic network.}
Here, the first term in the bracket is the change in the image caused by the change in $\mathbf{w}$, and $\Big|\Big|...\Big|\Big|_2$ denotes the 2-norm of the vector.  A constant difference $a$ in practice {helps stabilize the training process} \citep{Karras_2019}. {In practice, $a$ is obtained by the calculating the moving average of $\Big|\Big|\frac{\partial g(\mathbf{w})}{\partial \mathbf{w}}^{T}g(\mathbf{w})\Big|\Big|_2$ among the past 100 iterations.} Our final training objective is
\begin{equation}
    (G^*,D^*) = \arg \min\limits_{G} \max \limits_{D} \mathbb{E}_{\mathbf{z}\sim p(\mathbf{z}),\mathbf{x}\sim p(\mathbf{x})} (\mathcal{L}_{\rm adv}+\mathcal{L}_{r_1}+\mathcal{L_{\mathrm{path}}})
\end{equation}

\subsection{Few-shot Transfer Learning Technique}
\label{sec:fsgan}

\begin{figure*}
\centering
 \includegraphics[width=\linewidth]{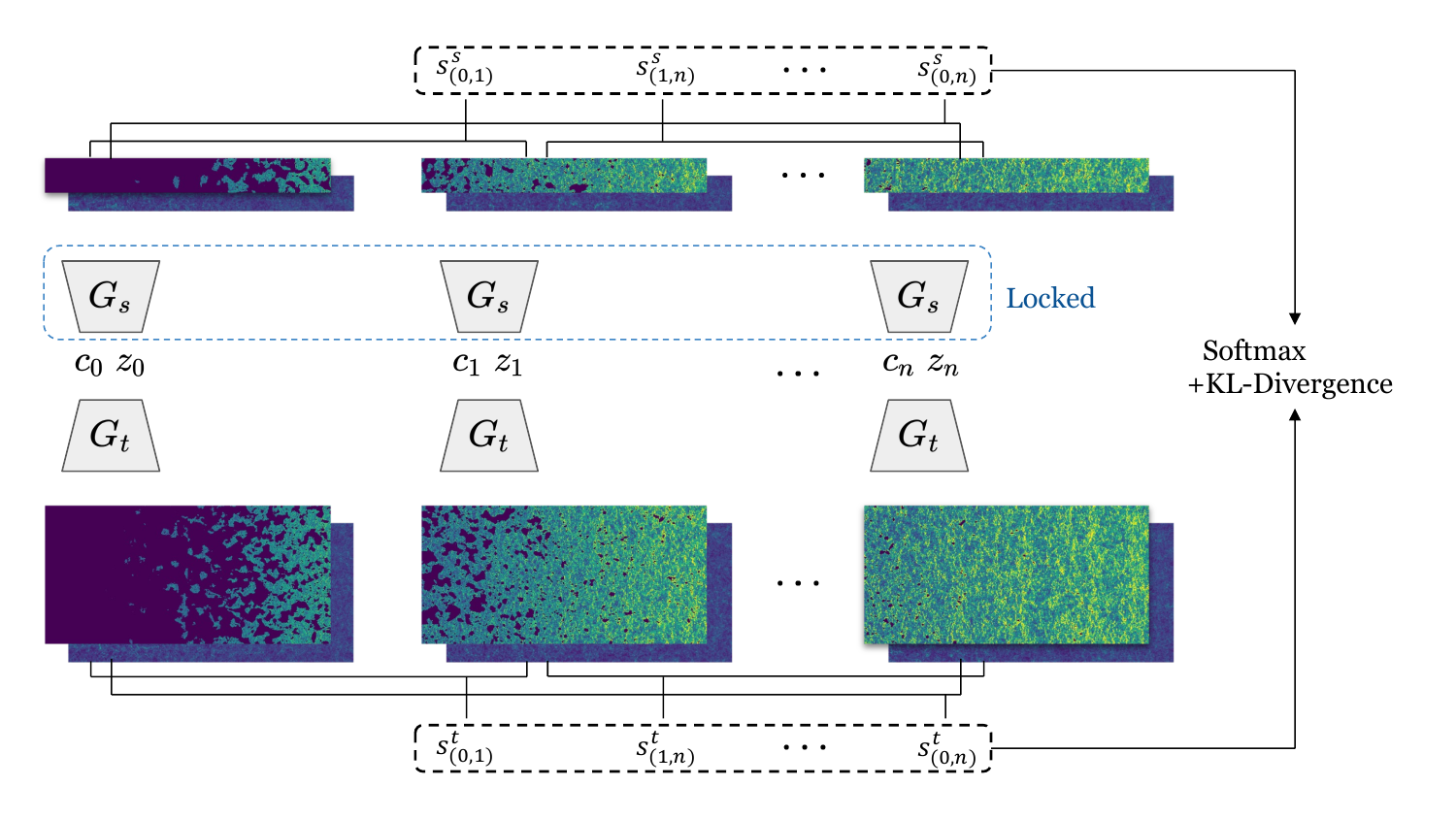}
 \caption{An illustration of CDC. We generate a set of samples with both small-scale GAN and large-scale GAN, calculate the similarity between each image pair generated by the same GAN, normalize the similarity vector of each GAN with softmax, and compute the KL-divergence as the CDC. }
 \label{fig:cdc}
\end{figure*}

Given a well-behaved small-scale \texttt{StyleGAN2} emulator, the network structure should be modified to enable the generation of large-scale images, before retraining it with our large-scale data set.  We adopt a simple approach where we expand the size of the generator's first layer, the \textsc{Constant Input} layer. {Suppose the size of the output layer is (C,H,W) (\{C, H, W\} stands for \{``channel'', ``height'', ``width''\}), the layer has a size of (C, H/$2^5$, W/$2^5$). } {We alter the shape of the layer} from $(2,2,16)$ to $(2,8,16)$ by duplicating the original layer four times, and concatenating them in the {height} axis. Consequently, after five rounds of upsampling {in spatial dimensions}, the final output size is $(2,256,512)$. 

We then retrain our GAN with large-scale images. We first employ the patchy-level discriminator and CDC as described in \citet{Ojha_2021}. We mark small-scale GAN as our source model $G_s$ and large-scale GAN as the target model $G_t$. We compute the CDC as follows. First, we use the same batch of vector $(\mathbf{z},\mathbf{c})$ feeding both $G_s$ and $G_t$, and get the corresponding small-scale image $G_s(\mathbf{z},\mathbf{c})$ and $G_t(\mathbf{z},\mathbf{c})$. Then, the set of computed similarity $\mathbf{S}_s(\mathbf{z},\mathbf{c})$ 
between any pair of images in the $G_s(\mathbf{z},\mathbf{c})$ sample is calculated as
\begin{equation}
    \mathbf{S}_s(\mathbf{z},\mathbf{c})=\{\cos(G_s(z_i,c_i),G_s(z_j,c_j))_{\forall i\neq j}\}\,,
\end{equation}
and the set of computed similarity $\mathbf{S}_t(\mathbf{z},\mathbf{c})$ from the  $G_t(\mathbf{z},\mathbf{c})$ sample is 
\begin{equation}
    \mathbf{S}_t(\mathbf{z},\mathbf{c})=\{\cos(G_t(z_i,c_i),G_t(z_j,c_j))_{\forall i\neq j}\}\,.
\end{equation}
Here ``$\cos$'' denotes the cosine similarity. 
Next, we normalize these two vectors in terms of softmax, 
\begin{equation}
    \text{softmax}(\mathbf{z})_i = \frac{e^{z_i}}{\sum_{j=1}^K e^{z_j}},
\end{equation}
where $\mathbf{z}$ is the vector to be normalized and $K$ is the length of $\mathbf{z}$. We further calculate the KL divergence between vectors
\begin{equation}
    \mathcal{L}_{\rm CDC} = D_{\rm KL}\left(\mathrm{softmax}(\mathbf{S}_s),\mathrm{softmax}(\mathbf{S}_t)\right)
\end{equation}
as the CDC loss. Figure \ref{fig:cdc} illustrates the idea of $\mathcal{L}_{\rm CDC}$. This treatment encourages $G_t$ to generate samples with a diversity similar to $G_s$, relieving the mode collapse problem.

In this work, a patchy-level discriminator is also adopted. Our training set consists only of 80 sets of parameters, which is not enough to cover the entire parameter space. Thus, we divided the whole parameter space into two parts: the anchor region and the rest. The anchor region is a spherical region around the training set parameters with a small radius. In this region, the GAN image $G_t(\mathbf{z},\mathbf{c}_{\rm anch})$ has a good training sample to compare with. Thus, we apply the full discriminator with these parameters. If $\mathbf{c}$ is located outside the anchor region, we apply only a patch discriminator. In this case, the discriminator does not calculate the loss of the whole image but calculates the loss of different patches of the image.  In practice, we sample from the anchor region and train $G_t$ with a fixed training epoch interval. This method reduces large-scale information usage and defers the happening time of the model collapse.

Since the small-scale information in both training sets is identical, we freeze the first two layers of the discriminator \citep{Mo_2020}, as they extract small-scale information that does not need modification. In addition, we add an extra adversarial loss term with small-scale discriminator $D_s$,
\begin{equation}
    \mathcal{L}_{\rm adv,s} ={\log(1- D_s(G_{t,{\rm cut}}(\mathbf{z},\mathbf{c})|\mathbf{c})) +\log (D_s(\mathbf{x}|\mathbf{c})})\,,
\end{equation}
to the loss function to ensure the robustness of small-scale information. $G_t(\mathbf{z},\mathbf{c})$ is cut into small pieces $G_{t,{\rm cut}}$ to fit the input size of $D_s$. Our implementation of these methods is publicly available in this GitHub repo\footnote{\url{https://github.com/dkn16/few-shot-gan-adaptation}}.

\section{Data preparation}
\label{sec:data}

In this section, we describe the process to generate our data set. We generate the EoR 21~cm mock signal using the semi-numerical simulation code \cmfst\ and create a training set that comprises of 30,000 small-scale simulations with a resolution of $(64,64,512)$ and 80 large-scale simulations with a resolution of $(256,256,512)$. The cell size for all simulations is $(2\,\rm Mpc)^{3}$.

\subsection{21cmFAST Simulation}
\label{sec:cmfst} 

The observable for the 21~cm line is the differential brightness temperature $T_b$ \citep[see, e.g.][]{Furlanetto2006,Mellema_2013}, 
\begin{equation}
    T_{b} \approx  27x_{{\rm HI}}(1+\delta_{m})\bigg(1-\frac{T_{\gamma}}{T_{{\rm S}}}\bigg) \bigg(\frac{1+z}{10}\frac{0.15}{\Omega_{m}h^{2}}\bigg)^{1/2}\bigg(\frac{\Omega_{b}h^{2}}{0.023}\bigg)
    \label{eqn:tb}
\end{equation}
in units of millikelvin. 
Here, $x_{\rm HI}$ is the neutral hydrogen fraction,  $\delta_m$ is the matter overdensity, $T_\gamma$ is the CMB temperature, and $T_{\rm S}$ is the spin temperature that characterizes the excitation status of hydrogen atoms between the hyperfine states. During EoR, the hydrogen gas was adequately heated, $T_{\rm S}\gg T_{\gamma}$. $\Omega_m$ and $\Omega_b$ are the matter density and baryon density with respect to the critical density in the current epoch, respectively.

\cmfst\ is a semi-numerical simulation code that uses the linear perturbation theory to yield initial condition, uses the second-order Lagrangian perturbation theory \citep[2LPT;][]{Scoccimarro_1998} to evolve the density field, and uses the excursion set theory \citep{Furlanetto_2004} to simulate the reionization process. Excursion set theory works by first generating the density field at a given redshift, then specifying the location of ionizing sources by introducing the minimum virial temperature $T_{\rm vir}$, the threshold virial temperature for a halo that can host ionizing sources. With $T_{\rm vir}$, sources will be assigned to high-density regions.  The parameter $\zeta$ is used to describe the number of photons emitted per baryon by an ionizing source. Together with the baryon collapsed fraction $f_{\rm coll}$, the total photons per unit volume emitted in a region are simply $n_{b}f_{\rm coll}\zeta$. One can then calculate a spherical region with a radius of $R$, to see if $\zeta f_{\rm coll}>1$, which is the criterion for a region to be fully ionized. To determine the largest possible value of $R$ that satisfies the criterion, an iteration from $R_{\rm mfp}$, the mean free path of ionizing photons, to the cell size $R_{\rm cell}$ is carried out for every source. The spherical region with this radius is then marked as fully ionized. If $R_{\rm cell}$ does not satisfy the criterion, the partial ionized fraction $x_{\rm HII}$ for this cell is set to $1/\zeta$. Finally, the differential 21~cm brightness temperature $T_b$ can be computed using Equation~(\ref{eqn:tb}).


Our reionization parameters are the ionizing efficiency $\zeta$ and the minimum virial temperature $T_{\rm vir}$.
\begin{itemize}
    \item \textit{Ionizing efficiency} $\zeta$.
    $\zeta$ is related to the number of ionizing photons emitted by an ionizing source and is defined as $\zeta = f_{\rm esc}f_{*}N_{\gamma}/(1+\bar{n}_{\rm rec})$, as a combination of several parameters which is still uncertain at high redshift \citep{Wise_2009}. Here, $f_{\rm esc}$ is the fraction of escaping photons from a galaxy into the IGM, $f_{*}$ is the fraction of baryons that collapsed into the stars in the galaxy, $N_{\gamma}$ is the number of photons produced per baryon in the star, and $\bar{n}_{\rm rec}$ is the mean recombination rate per baryon. We explored a range of $1<\log_{10}\zeta<2.398$.
    
    \item \textit{Minimum virial temperature} $T_{\rm vir}$. $T_{\rm vir}$ corresponds to the minimum mass of haloes that can host ionizing sources. This parameter implies the underlying physics of star and galaxy formation in dark matter haloes. In our data set, we set the range of $T_{\rm vir}$ as $4<\log_{10} T_{ \rm vir}<6$.
\end{itemize}

\subsection{Training Data Set}

Our data set consists of two parts --- a small-scale data set and a large-scale data set. {We choose flat priors for both parameters, $\log \zeta \sim \mathcal{U}[1,2.398]$ and $\log T_{\rm vir}\sim \mathcal{U}[4,6]$, where $\mathcal{U}[a,b]$ represents flat prior from $a$ to $b$.}

The small-scale set has a resolution of $(64,64,512)$. The data set for training the small-scale GAN consists of 30,000 lightcone simulations with a comoving length of $(128,128,1024)$ Mpc. The third axis ($z$-axis) is along the line of sight (LoS), spanning a redshift range of $7.51<z<11.93$. 
Each such lightcone simulation box is concatenated by eight cubic boxes (each with different initial conditions). For each box, we run a simulation realization with a grid resolution of $64^3$ in a comoving volume of $(128\,{\rm Mpc})^3$. For each redshift, we pick up the image slice at the corresponding comoving position in the corresponding cubic box at the corresponding cosmic time and load it into the lightcone simulation box. In other words, every 64 slices are in the same realization. Each lightcone simulation box in our small-scale data set takes 0.3 core hours and a memory of 1.2 GB. We include both the overdensity field and 21~cm $T_b$ field for training. For each lightcone simulation box, we cut two slices in the $x$-axis and two slices in the $y$-axis {with a separation of 64 Mpc between slices to minimize the similarities between slices, to avoid the mode collapse due to the clustering of similar slices. This cut} results in 120,000 lightcone images with the size of $(2,64,512)$ in our small-scale data set, where the first channel is the $T_b$ field and the second channel is the $\delta_m$ overdensity field. {We choose the size of the small-scale set to be 120,000 because GAN typically requires $\gtrsim10^5$ images as the training set \citep[e.g.][]{2020arXiv200606676K}.}

The large-scale set has a resolution of $(256,256,512)$. The data set for training the large-scale GAN consists of 80 lightcone simulations with a comoving length of $(512,512,1024)$ Mpc. The third axis covers the same redshift range along the LoS as in the small-scale set. 
Each such lightcone simulation box is concatenated by two cubic boxes (each with different initial conditions). For each box, we run a simulation realization with a grid resolution of $256^3$ in a comoving volume of $(512\,{\rm Mpc})^3$. We synthesize the lightcone simulation box from the realizations of cubic boxes in the similar approach to the small-scale set. Every 256 slices are in the same realization. 
Each lightcone simulation box in our large-scale data set takes 30 core hours and a memory of 29 GB. For each lightcone simulation box, we cut two slices on the $x$-axis and two slices on the $y$-axis, resulting in 320 lightcone images with the size of $(2,256,512)$ in our large-scale data set, containing both brightness temperature field and overdensity field for training.

For the training set, each lightcone simulation has a different set of reionization parameters. In both small-scale and large-scale training sets, we use the Latin Hypercube Sampling method \citep{Mckay_2000} to sample the parameters, because this method ensures the homogeneity of parameter sample distribution in the parameter space. Therefore, while 30,000 lightcone simulations in the small-scale training set cover a wide range of parameter space, 80 lightcone simulations in the large-scale training set can also act as a representative set of parameters. 

\subsection{Test Data Set}

For the test sets of both small-scale and large-scale GAN, we choose the same five sets of parameters that are equal-space sampled in the parameter space: $(\log_{10}\zeta,\log_{10} T_{\rm vir})=(1.35,5.50),\ (1.7,5.0),\ (2.05,4.5),\ (1.35, 4.50)$ and $(2.05,5.5)$. {The ionization histories for the test sets span a wide range (e.g.\ ending at $z\sim 6 - 9$), and fully cover current models and constraints \citep[e.g. Figure 9 in ][]{2024arXiv241209732F}.}
For illustration purposes, we will show the results only for the first three cases in the rest of this paper, but the conclusions made herein are generic and based on the tests with all five sets. 

For the test set of small-scale GAN, we run 100 realizations of lightcone simulation boxes for each parameter set to minimize the impact of cosmic variance. 
For each realization, we extract 64 slices of lightcone images. So, the test set for evaluating the small-scale GAN consists of 6,400 image samples for each parameter set, or a total of 32,000 image samples for five parameter sets in a total of 500 realizations. 

However, for the test set of large-scale GAN, limited by computational resources, we run only four realizations of lightcone simulation boxes for each parameter set. For each realization, we extract 256 slices of lightcone images. So, the test set for evaluating the large-scale GAN consists of 1,024 image samples for each parameter set, or a total of 5,120 image samples for five parameter sets in a total of 20 realizations.

\section{Small-scale GAN Preparation}
\label{sec:ssgan}
\begin{figure*}
\centering
 \includegraphics[width=0.75\linewidth]{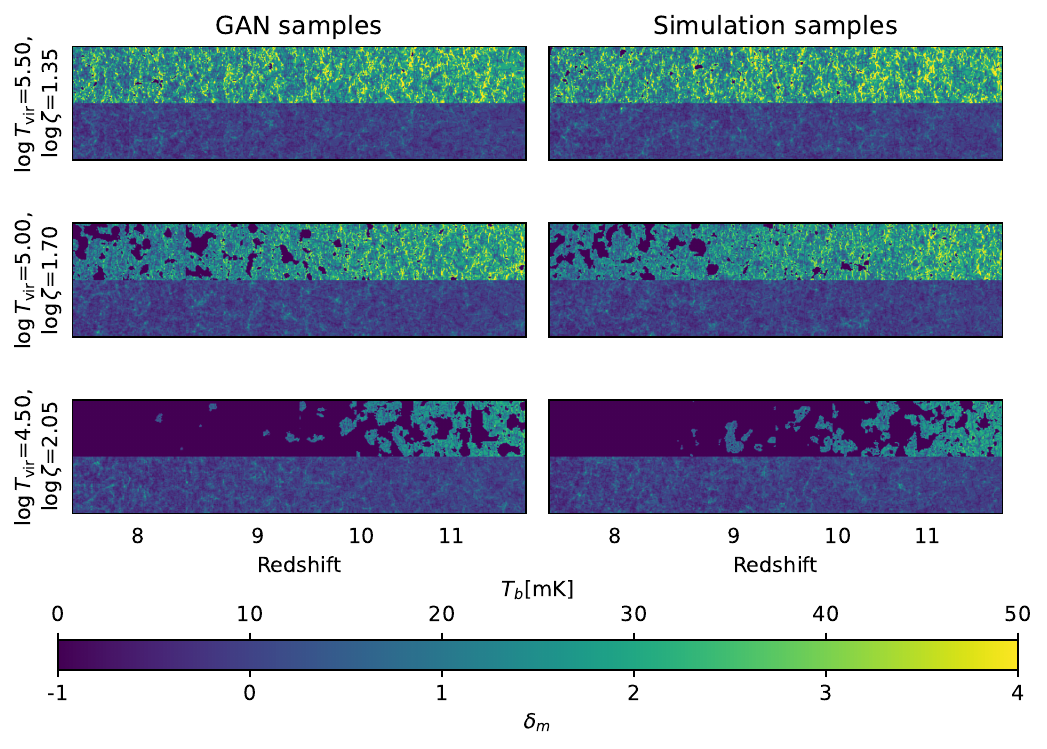}
 \caption{Examples of the emulated images using the small-scale GAN (left), in comparison with the simulated images using 21cmFAST in the test set (right). In each panel, we show the 21~cm brightness temperature ($T_b$) field (the upper half) and the matter overdensity ($\delta_m$) field (the lower half). The LoS is along the x-axis.}
 \label{fig:samp64}
\end{figure*}
Our first step is training a small-scale GAN with sufficient data (i.e.\ 120,000 image samples from 30,000 simulations in this paper). Our training configurations are discussed in Appendix \ref{app:trainconf}. Since our training process is a two-step process, it is necessary to evaluate the output of the small-scale GAN.
The test set for evaluating the small-scale GAN consists of 32,000 image samples for five parameter sets in a total of 500 realizations. 

A visual comparison of our test samples, as shown in Figure~\ref{fig:samp64}, implies that our model reproduces the features of ionized bubbles in the $T_b$ field and the cosmic web structures in the $\delta_m$ field. The ionized bubble size evolution is clearly visible in the GAN samples, and the density field shows a clearer web structure as redshift evolves. Furthermore, the bubble size and number density vary with reionization parameters.

\subsection{Global Signal}
\begin{figure*}
\centering
 \includegraphics[width=0.8\linewidth]{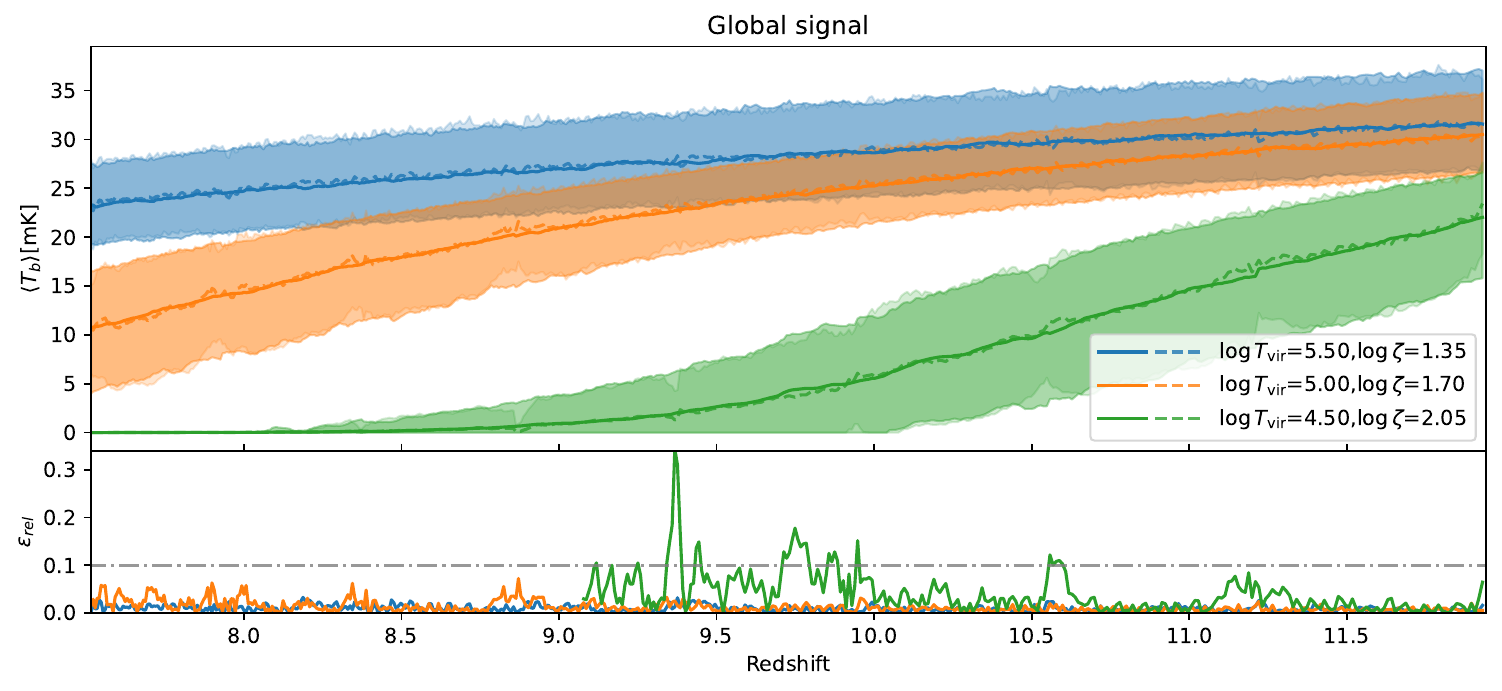}
 \caption{(Top) the global 21~cm signal emulated with the small-scale GAN, with the mean (dashed line) and $2\sigma$ scatter (shallow shaded region). We show the results with different values of reionization parameters (in different colors). Each set of reionization parameter is calculated with 6,400 image samples. For comparison, we show the results of test set images, with the mean (solid line) and $2\sigma$ scatter (thick shaded region). (Bottom) relative error between the GAN-emulated global signal and the test set. For visualization, the 10\% error level is indicated with the dot-dashed line, while neglecting the data points at which the statistics for the test set are nearly zero.}
 \label{fig:glob64}
\end{figure*}
Figure~\ref{fig:glob64} presents the global $T_b$ signal emulated with the small-scale GAN under different sets of parameters, with each parameter set calculated with 6,400 samples. The relative error $\varepsilon_{\rm rel}$ is defined as the average of the statistics evaluated by the GAN divided by the average of the statistics evaluated by the test set, minus 1:
\begin{equation}
    \varepsilon_{\rm rel}{({\rm Stats})} = \frac{\left<\mathrm{Stats}_{\mathrm{GAN}}\right>}{\left<\mathrm{Stats}_{\mathrm{test}}\right>}-1 \label{eq:relerr}
\end{equation}
Here, ``$\rm Stats$'' denotes the statistics we choose to evaluate the GAN; in this case, it is the global signal. A cutoff is performed when $\mathrm{Stats_{test}}$ is close to zero. We present three sets of parameters in this figure, each with a unique reionization history. Our GAN samples accurately reproduce the global signal, as seen in the relative error plot. We find that when $\left<T_b\right>$ is large, the relative error can be at the subpercent level, but when $\left<T_b\right>$ is small (so is the demoninator in Equation~\ref{eq:relerr}), the error can reach the level of tens of percent.  The $2 \sigma$ scatter of the GAN result and the test set demonstrates a good agreement with each other for different sets of parameters, which implies that the GAN might be extensively applicable to a wide range of reionization parameters. 
\subsection{Power Spectrum}
\begin{figure*}
\begin{center}
\begin{tabular}{ccc}
\includegraphics[width=0.3\linewidth]{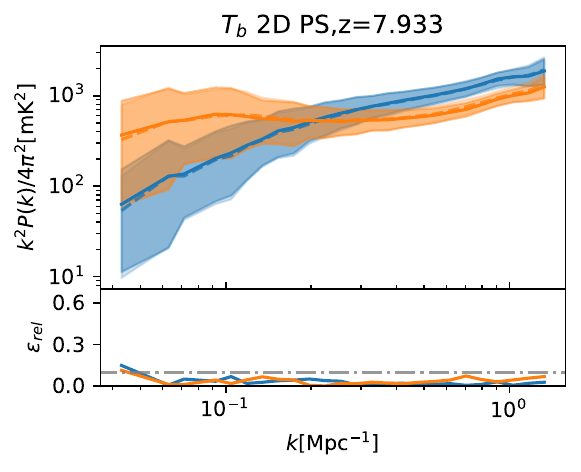}
\includegraphics[width=0.3\linewidth]{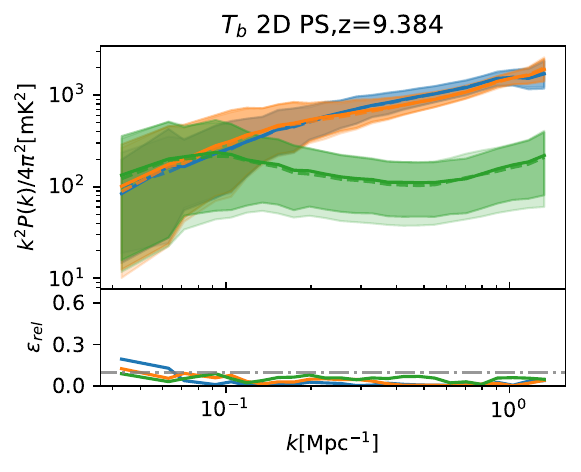}
\includegraphics[width=0.3\linewidth]{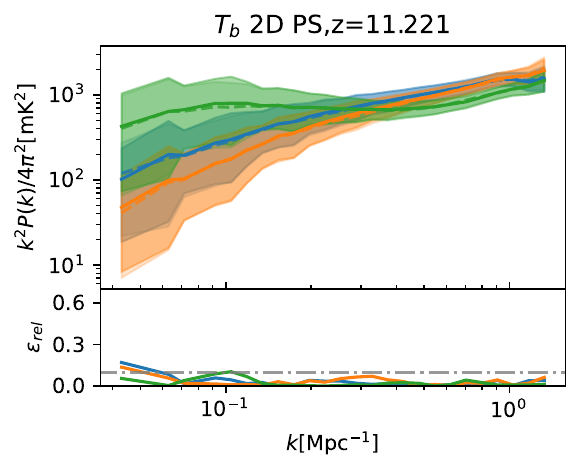}\\
\includegraphics[width=0.3\linewidth]{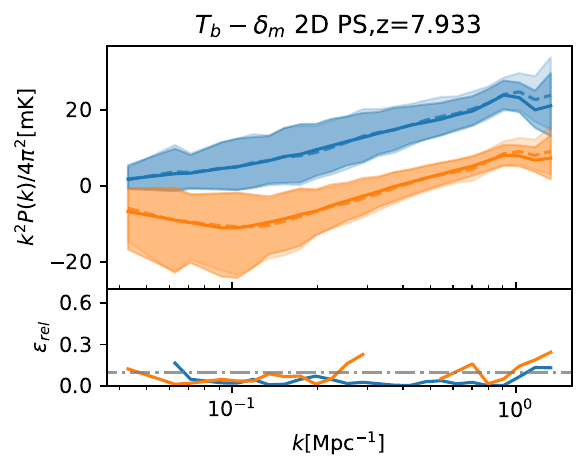}
\includegraphics[width=0.3\linewidth]{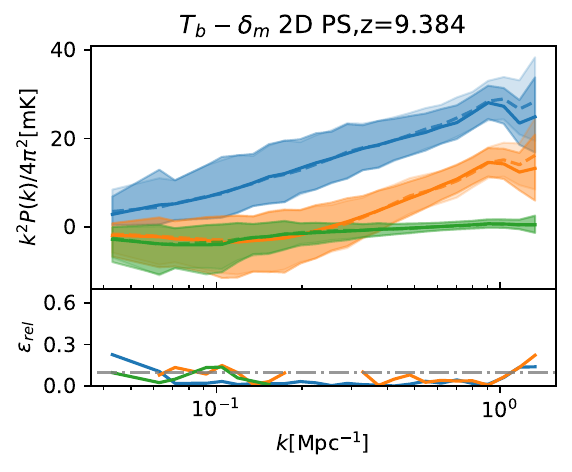}
\includegraphics[width=0.3\linewidth]{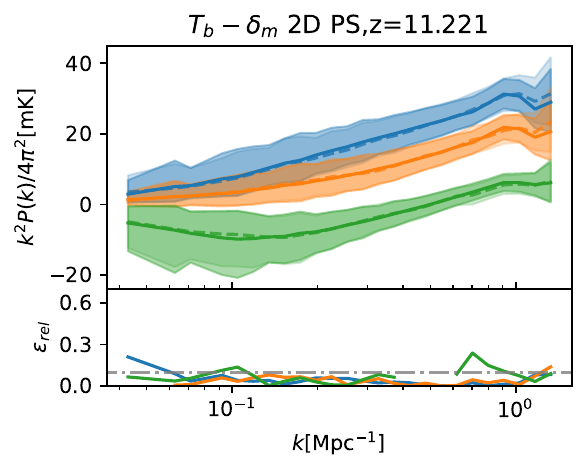}\\
\includegraphics[width=0.3\linewidth]{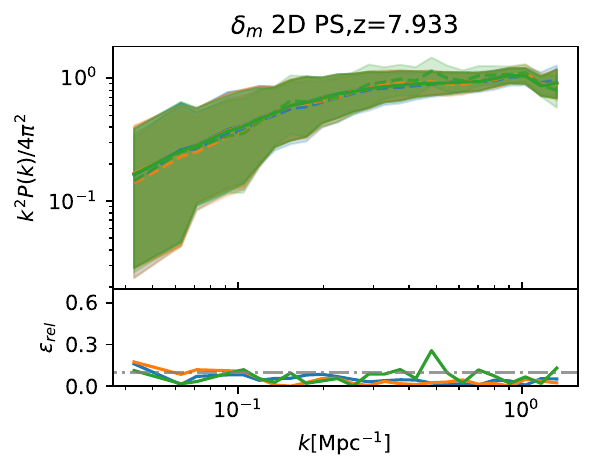}
\includegraphics[width=0.3\linewidth]{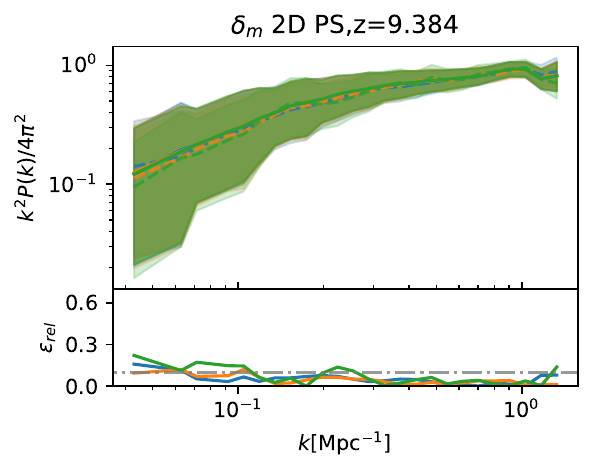}
\includegraphics[width=0.3\linewidth]{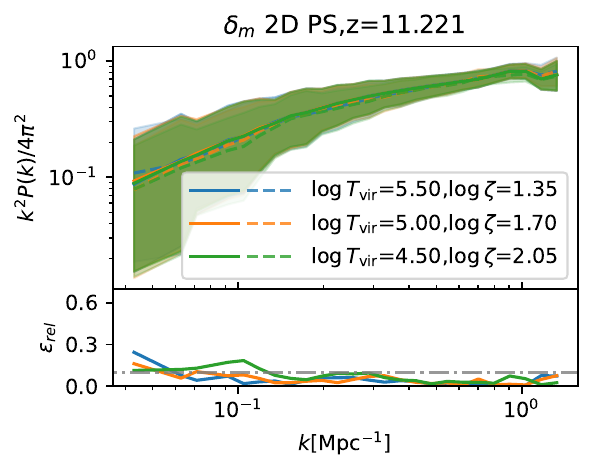}\\
\end{tabular}
\end{center}
\caption{The 2D power spectrum of the 21~cm signal emulated with the small-scale GAN with the mean (dashed line) and $2\sigma$ scatter (shallow shaded region). We show the results with different values of reionization parameters (in different colors). Each set of reionization parameter is calculated with 6,400 clips of size (2,64,64) from the raw image. For comparison, we show the results of test set images, with the mean (solid line) and $2\sigma$ scatter (thick shaded region). From top to bottom, we show the auto-PS of the 21~cm field $T_{b}$, the cross-PS between $T_{b}$ and the matter overdensity field $\delta_m$, and the auto-PS of $\delta_m$, respectively, at three representative redshifts of the center slice (from left to right) $z=7.933$, $9.384$, and $11.221$. The lower sub-panel in each panel shows the relative error between the GAN-emulated PS and the test set, $\varepsilon_{\rm rel}$. For visualization, the 10\% error level is indicated with the grey dot-dashed line, while neglecting the data points at which the statistics for the test set are nearly zero.  
Note that the case of $(\log_{10}\zeta = 2.05 ,\log_{10} T_{\rm vir}=4.5)$ (green) is completely ionized at $z=7.933$, so the 2D PS is not shown in the top left and middle left panels.}
\label{Fig:PS64}
\end{figure*}

One of the most commonly studied statistics in EoR is the power spectrum (PS). In Figure~\ref{Fig:PS64}, we show a comparison of the 2D PS between the GAN results and the test set with three sets of EoR parameters. For each parameter set, we use 6,400 GAN samples and 6,400 test samples to calculate the statistics. 

We find that, not only do the mean values of the GAN results and the test set agree well with each other, but also the $2\sigma$ scatter regions overlap in each plot. This agreement is observed for different sets of reionization parameters, which again supports the diversity of the GAN samples. 

Our GAN performs well in recovering the correlation between fields, which is not a simple task for GAN especially when two fields have different dependencies on parameters. We find that at all stages of EoR, the relative errors of the $T_b$ auto-PS, the $\delta_m$ auto-PS, and the $T_{b}$-$\delta_m$ cross-PS are mostly at the percent level, and overall below 20\% even when the value of PS is small.


\subsection{Non-Gaussianity}

\begin{figure}
\begin{center}
\begin{tabular}{c}
\includegraphics[width=0.85\linewidth]{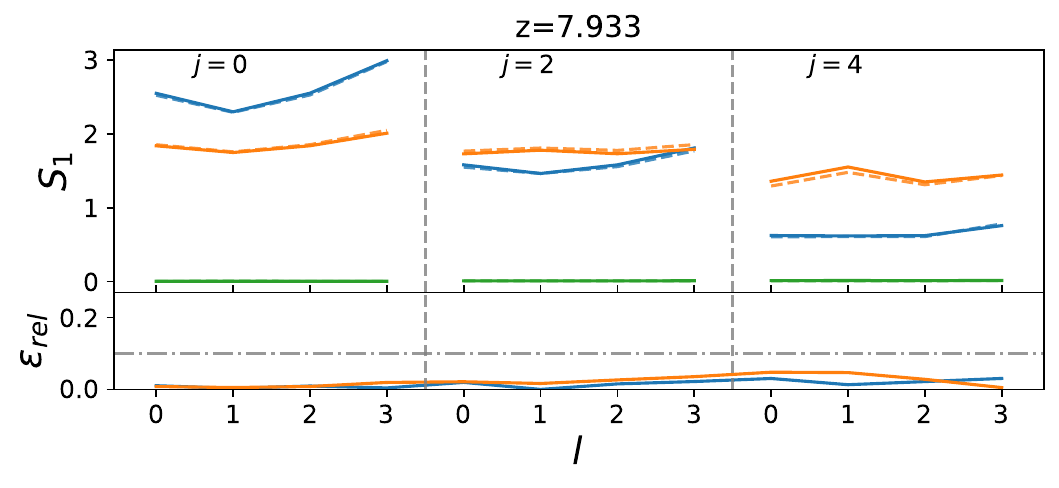}\\
\includegraphics[width=0.85\linewidth]{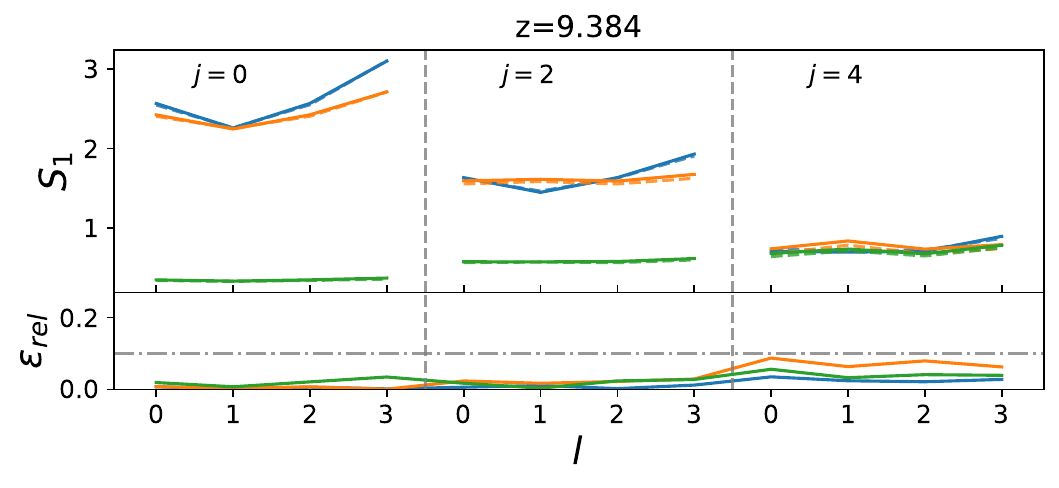}\\
\includegraphics[width=0.85\linewidth]{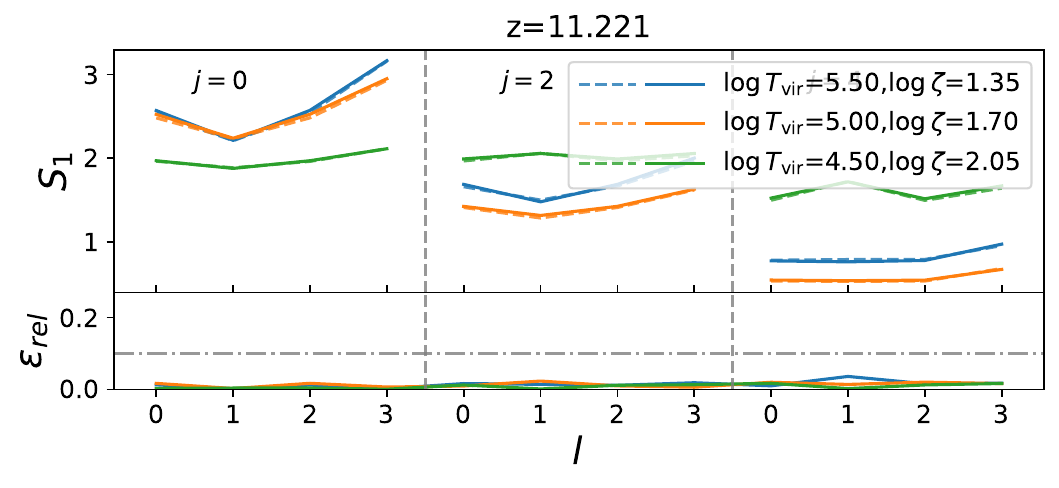}
\end{tabular}
\end{center}
\caption{The ST coefficient $S_{1}(j,l)$ of the 21~cm signal emulated with the small-scale GAN (dashed line) and that of the test set images (solid line) at three representative redshifts of the center slice (from top to bottom) $z=7.933$, $9.384$, and $11.221$, respectively. We show the results with different values of reionization parameters (in different colors). Each set of reionization parameter is calculated with 6,400 clips of size (2,64,64) from the raw image. The lower sub-panel in each panel shows the relative error between the GAN-emulated ST and the test set, $\varepsilon_{\rm rel}$. For visualization, the 10\% error level is indicated with the grey dot-dashed line, while neglecting the data points at which the statistics for the test set are nearly zero. }
\label{Fig:S164}
\end{figure}

\begin{figure}
\begin{center}
\begin{tabular}{c}
\includegraphics[width=0.85\linewidth]{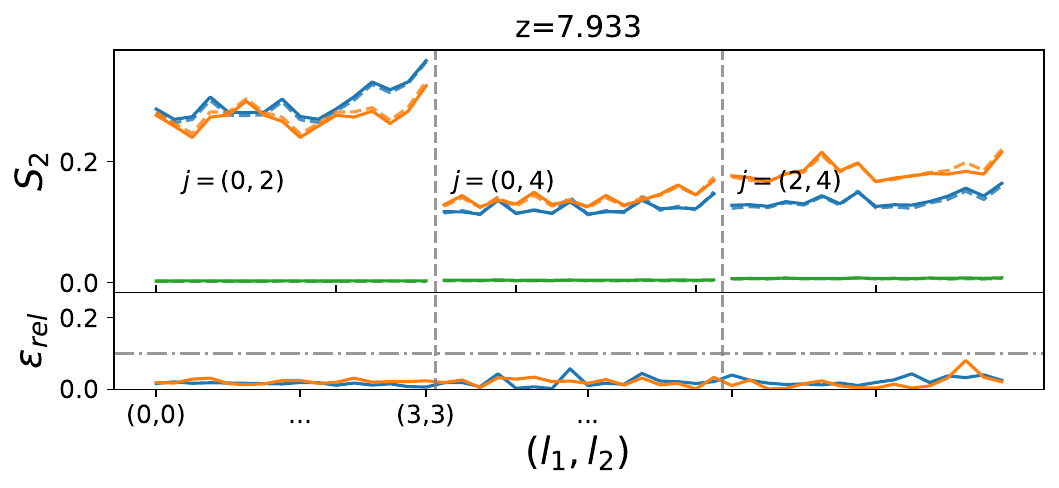}\\
\includegraphics[width=0.85\linewidth]{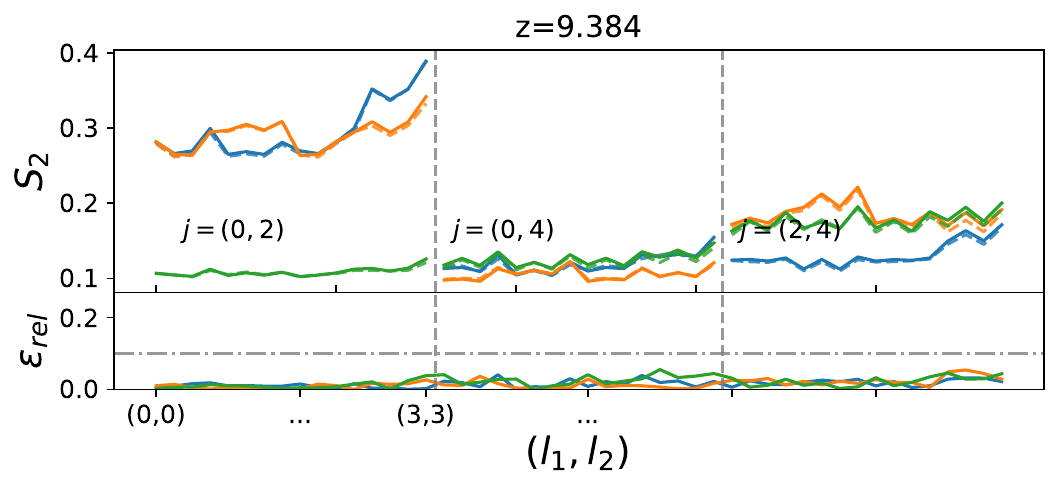}\\
\includegraphics[width=0.85\linewidth]{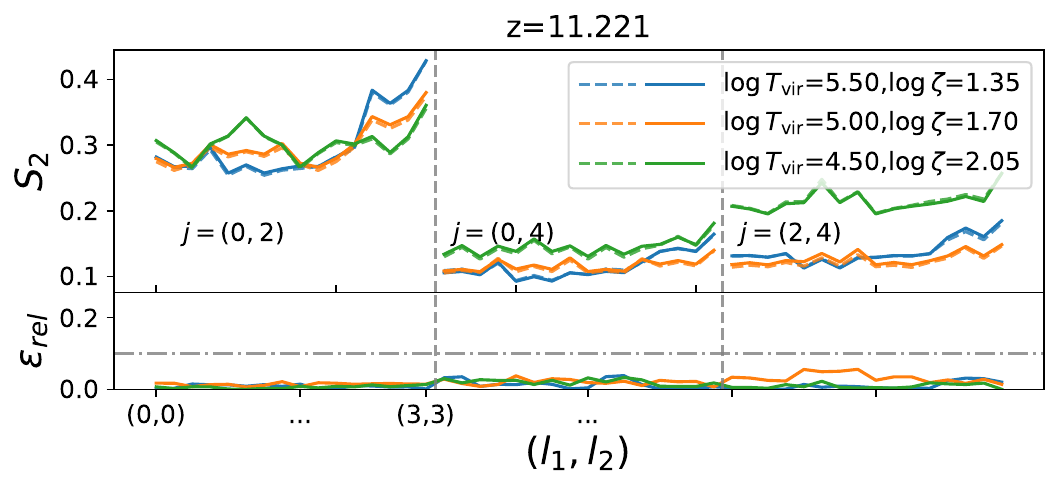}
\end{tabular}
\end{center}
\caption{Same as Figure~\ref{Fig:S164} but for $S_{2}(j_1,l_1,j_2,l_2)$. The indices $(j_1,j_2)$ take the value of $(0,2)$, $(0,4)$ and $(2,4)$. For each combination of $(j_1,j_2)$, the indices $(l_1,l_2)$ run from $l_1=0,1,2,3$ and $l_2=0,1,2,3$, i.e.\ $(l_1,l_2)$ take 16 values of $(0,0),(0,1),(0,2),(0,3),(1,0),(1,1)\ldots (3,3)$, respectively.}
\label{Fig:S264}
\end{figure}

To capture the non-Gaussian feature beyond the PS, we employ the scattering transform \citep[ST; e.g.][]{Mallat_2012,Allys_2019,Cheng_2020,Greig_2022} as a non-Gaussian statistic to evaluate our GAN. We refer interested readers to \citet{Cheng_2021} for a detailed description of ST. 

The ST coefficients $S_1$ and $S_2$ are defined as
\begin{equation}
    \begin{aligned}
        I_{1}(j,l) &= \left|I_0\ast\Psi\left(j,l\right)\right|\ast{\Phi(j)}\\
        I_{2}(j_1,l_1,j_2,l_2) &=\left| \left|I_0\ast\Psi\left(j_1,l_1\right)\right|\ast\Psi\left(j_2,l_2\right)\right|\ast {\Phi(j_2)}\\
        S_{1}(j,l)& =  \left<I_{1}(j,l)\right>\\
        S_{2}(j_1,l_1,j_2,l_2)& = \left<I_{2}(j_1,l_1,j_2,l_2)\right>
    \end{aligned}
\end{equation}
Here, $I_0$ is the input field. {In our work, $I_0$ is the 21 cm $T_b$ field. We leave out the density field because it is highly Gaussian. ``$\ast$'' denotes the convolution,} $\Psi$ is the {Morlet} wavelet kernel (see e.g. Appendix~B of \citealt{Cheng_2020} for its definition). The index $j$ defines the scale of the convolutional kernel --- the smaller $j$ corresponds to a more local kernel. The index $l$ defines the orientation of the kernel. {$\Phi(j)$ is the 2D Gaussian kernel with the same standard deviation $\sigma = 0.8\times2^{j-1}$ in both spatial dimensions to smear out the small-scale fluctuations in $\{I_1,I_2\}$.}
Here we choose $j = {0,2,4}$ and $l=0,1,2,3$ to cover a wide range of scales and orientations, resulting in 12 coefficients for $S_{1}(j,l)$ and 48 coefficients for $S_{2}(j_1,l_1,j_2,l_2)$. The ST coefficients are calculated using \textsc{Kymatio}\footnote{\url{https://github.com/kymatio/kymatio}} \citep{Andreux_2020}.

We show the comparison of ST coefficients of the small-scale GAN and the test set, averaged over 6,400 samples for each parameter set, in Figure~\ref{Fig:S164} (for $S_1$) and Figure~\ref{Fig:S264} (for $S_2$), respectively. The GAN results show very good agreement with the test set. The relative error is mostly below $5\%$, and overall below $10\%$. {We also compare the relative error of mean value and $2\sigma$ scatter in Table \ref{tab:rel_err}. Here the relative error of $2\sigma$ scatter is the average of the relative error of the $2.5\%$ and $97.5\%$ percentile, respectively, of the ST coefficient between GAN samples and test samples.  We find that the accuracies of mean value and $2\sigma$ scatter are at the same level, which is an indication of no strong mode collapse.} 

In sum, our small-scale GAN trained with 120,000 image samples that represent 30,000 sets of reionization parameters is shown to emulate the \cmfst\ simulation with high precision in both mean value and statistical scatter of several statistics. 
The small-scale GAN, therefore, serves as an excellent starting point for our second step, i.e.\ training the large-scale GAN.

\section{Result: Large-scale GAN}
\label{sec:lsgan}
\begin{figure*}
\centering
 \includegraphics[width=0.7\linewidth]{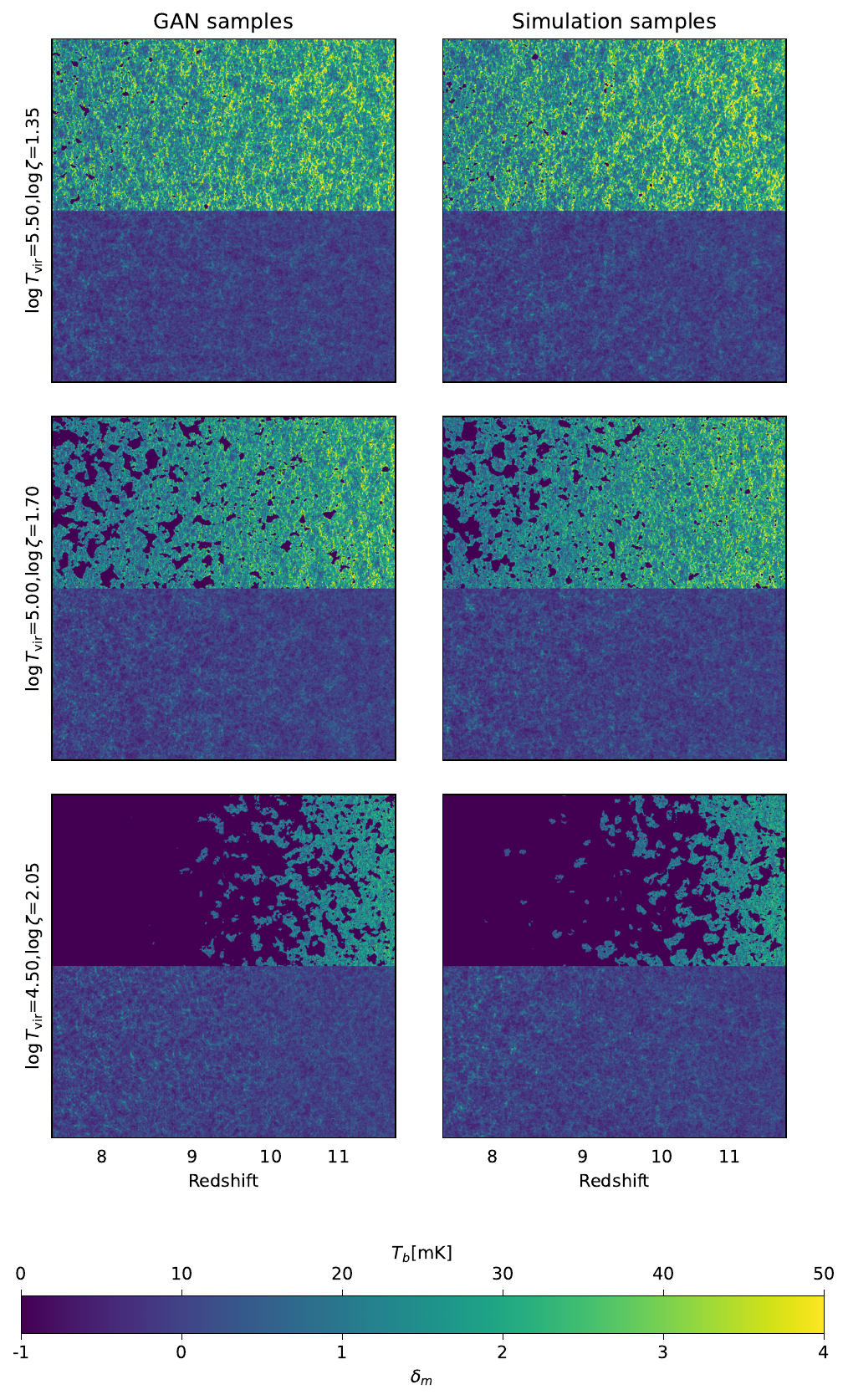}
 \caption{Same as Figure~\ref{fig:samp64} but for the large-scale GAN.}
 \label{fig:samp256}
\end{figure*}

Our final objective is training the large-scale GAN with a limited data set. To do so, we apply the few-shot transfer learning techniques described in Section~\ref{sec:fsgan} and generate a large-scale GAN using the small-scale GAN that was trained and tested in Section~\ref{sec:ssgan}. We train the large-scale GAN for {1,400 epochs}, using a training set that consists of only 320 image samples from 80 simulations. The test set for evaluating the large-scale GAN consists of 5,120 image samples for five parameter sets in a total of 20 realizations. 

A visual inspection of the test samples of the large-scale GAN is shown in Figure~\ref{fig:samp256}. We find that the concatenating boundaries in the test samples of the small-scale GAN now disappear in the test samples of the large-scale GAN due to retraining, an evidence of improved image quality by our GAN. {Moreover, This significantly outperforms the GAN trained only with 80 large-scale simulations, as presented in Appendix~\ref{app:80}.}

\subsection{Global Signal}
\begin{figure*}
\centering
 \includegraphics[width=0.9\linewidth]{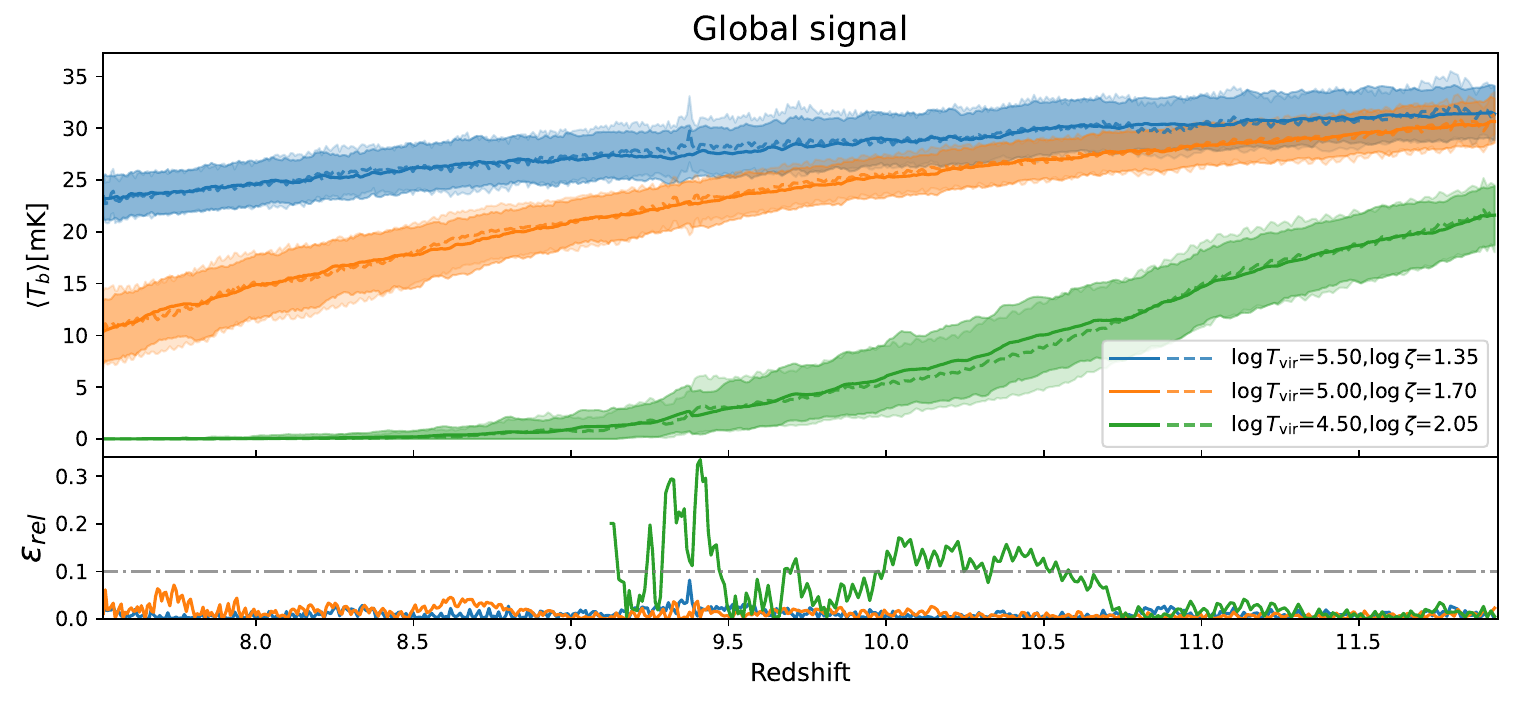}
 \caption{Same as Figure~\ref{fig:glob64} but for the large-scale GAN. Each set of reionization parameter is calculated with 1,024 image samples.}
 \label{fig:glob256}
\end{figure*}

Figure~\ref{fig:glob256} presents the global $T_b$ signal emulated with the large-scale GAN. Limited by the size of test set, the mean value is calculated with 1,024 image samples for each parameter set. The large-scale GAN results are slightly worse than the small-scale GAN. For example, for the case of $(\log_{10}\zeta = 2.05 ,\log_{10} T_{\rm vir}=4.5)$, the relative error exceeds $10\%$ at the early stage of reionization, and the $2\sigma$ scatter is also slightly larger than the test set. However, for the other two cases of reionization parameters, the large-scale GAN still performs well, with an error of less than 5\% and a well-matched $2\sigma$ scatter region. 
\subsection{Power Spectrum}
\begin{figure*}
\begin{center}
\begin{tabular}{ccc}
\includegraphics[width=0.3\linewidth]{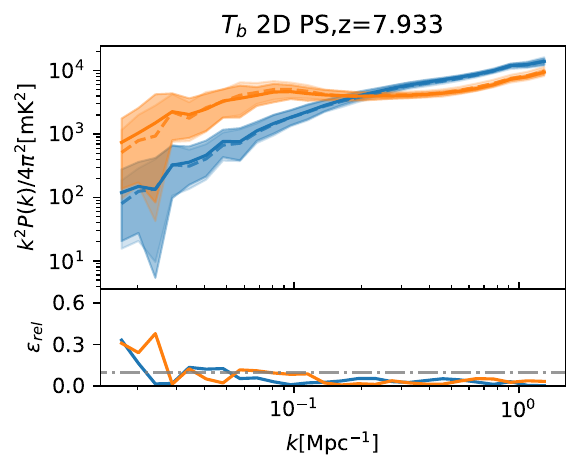}
\includegraphics[width=0.3\linewidth]{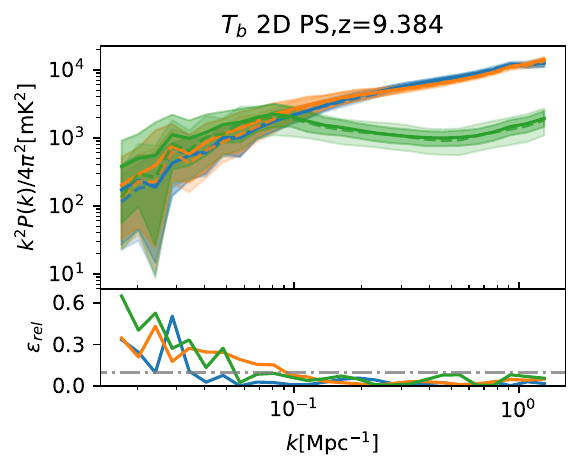}
\includegraphics[width=0.3\linewidth]{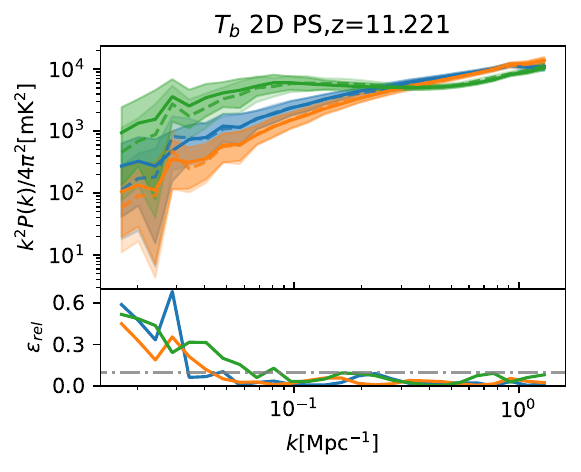}\\
\includegraphics[width=0.3\linewidth]{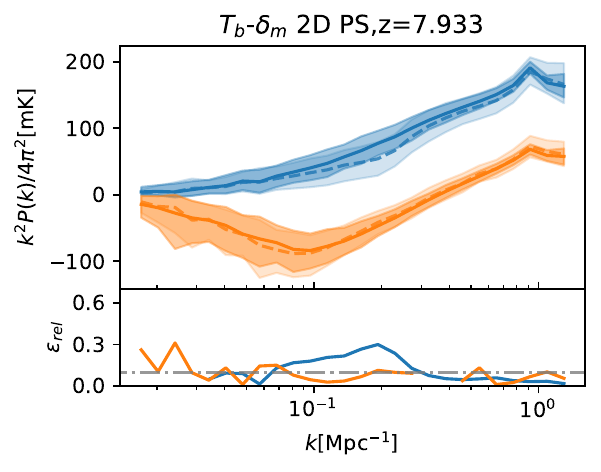}
\includegraphics[width=0.3\linewidth]{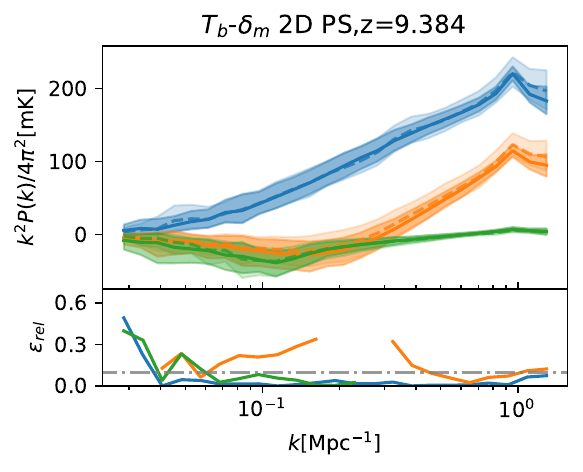}
\includegraphics[width=0.3\linewidth]{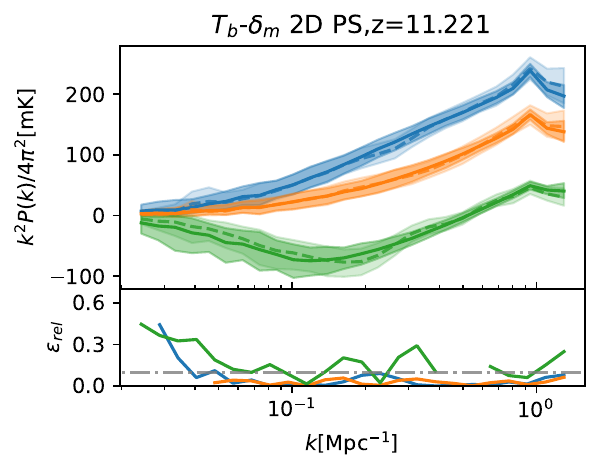}\\
\includegraphics[width=0.3\linewidth]{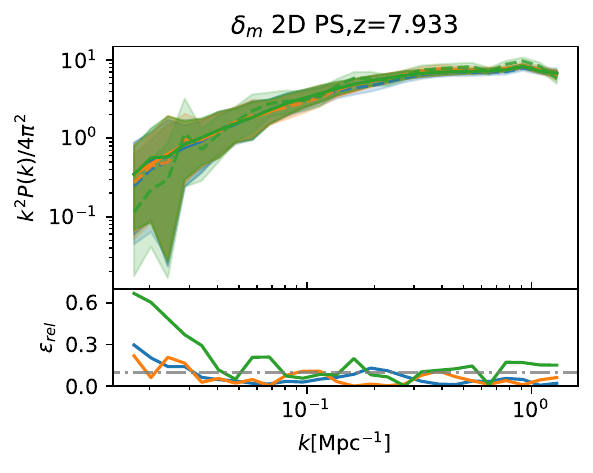}
\includegraphics[width=0.3\linewidth]{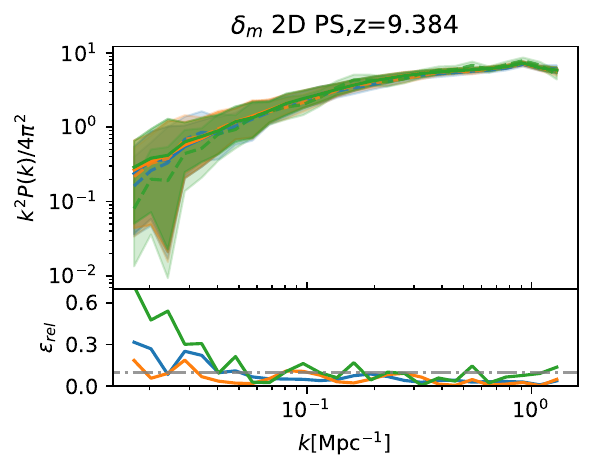}
\includegraphics[width=0.3\linewidth]{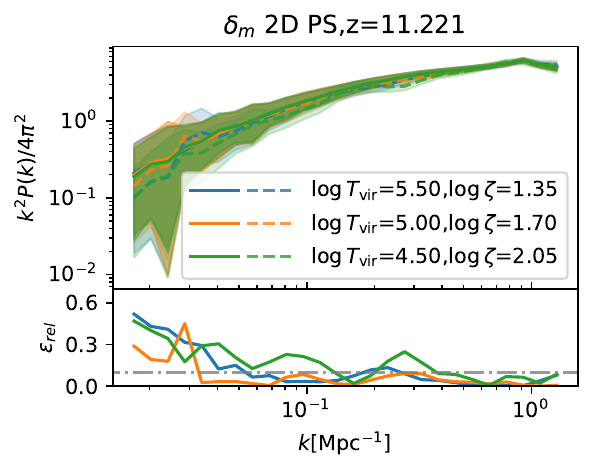}\\
\end{tabular}
\end{center}
\caption{Same as Figure~\ref{Fig:PS64} but for the large-scale GAN. Each set of reionization parameter is calculated with 1,024 clips of size $(2,256,128)$ from the raw image. } 
\label{Fig:PS256}
\end{figure*}

In Figure~\ref{Fig:PS256}, we show a comparison of 2D PS between the large-scale GAN results and the test set.  
The GAN performs well on small scales, with relative error below $10\%$. However, on very large scales ($k\lesssim 0.02\,{\rm Mpc}^{-1}$), the relative error can be $\gtrsim 30\%$. This result is not surprising because the features on large scales have not been well trained due to very limited large-scale training sets, but given that the training set for large-scale GAN is only 320 lightcone images, this level of error is acceptable.

The $2\sigma$ scatter of the PS for the large-scale test set is much smaller than that for the small-scale test set, because more modes are included within an image sample. A similar trend is found in the GAN results, i.e.\ the sampling variance for the large-scale GAN is much less than that for the small-scale GAN. 

\subsection{Non-Gaussianity}
\begin{figure}
\begin{center}
\begin{tabular}{c}
\includegraphics[width=0.85\linewidth]{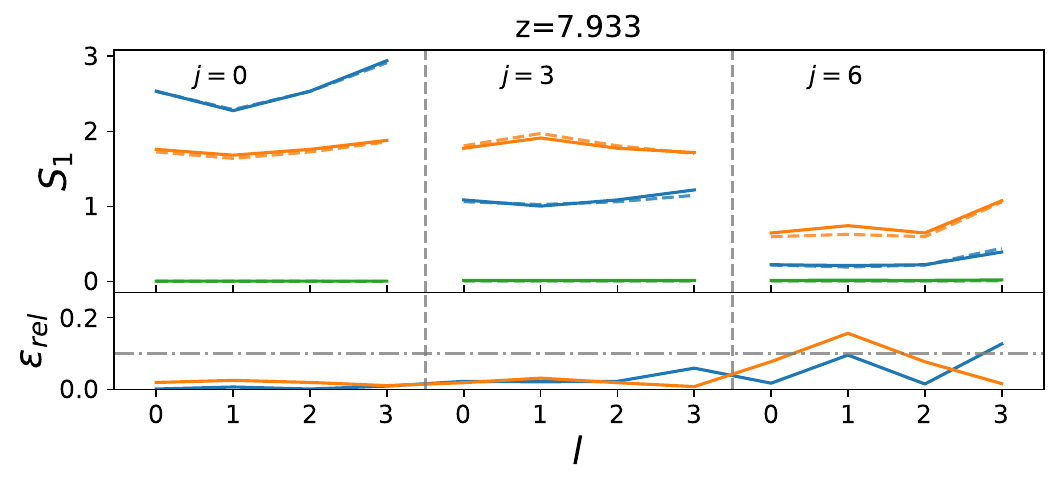}\\
\includegraphics[width=0.85\linewidth]{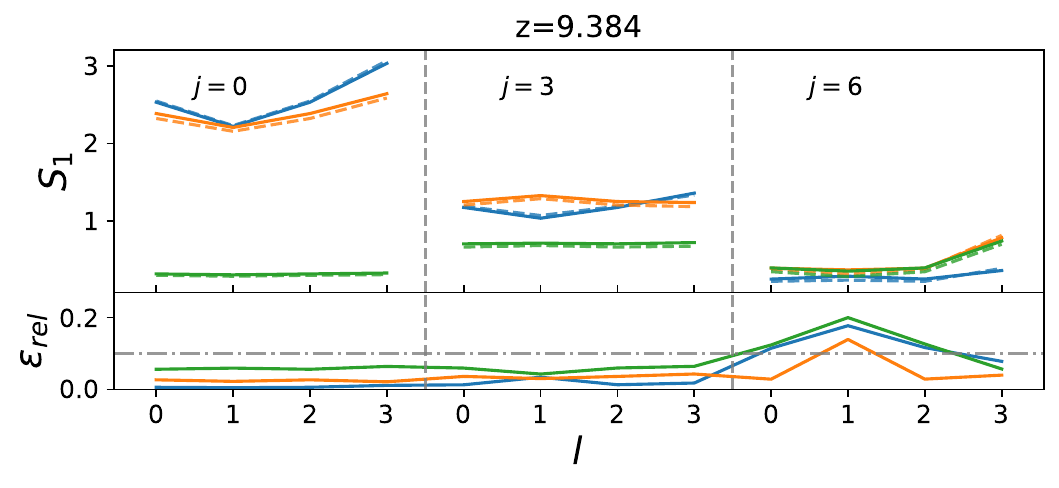}\\
\includegraphics[width=0.85\linewidth]{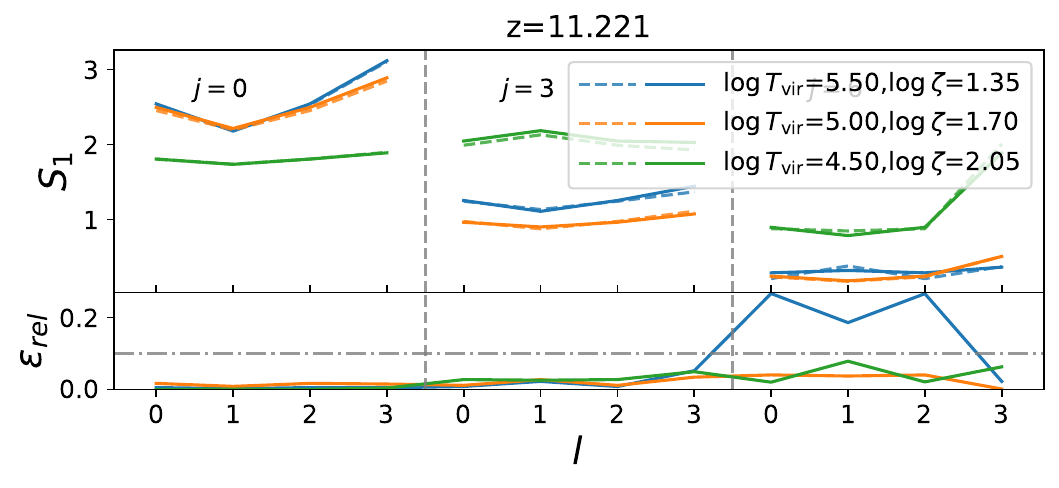}
\end{tabular}
\end{center}
\caption{Same as Figure~\ref{Fig:S164} but for the large-scale GAN and a choice of the index $j=0,3,6$. Each set of reionization parameter is calculated with 1,024 clips of size $(2,256,128)$ from the raw image.}
\label{Fig:S1256}
\end{figure}

\begin{figure}
\begin{center}
\begin{tabular}{c}
\includegraphics[width=0.85\linewidth]{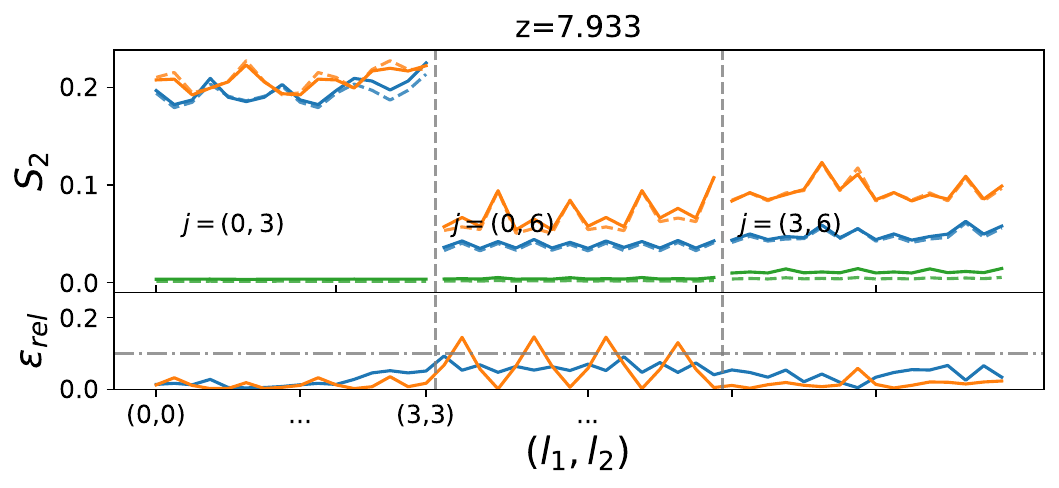}\\
\includegraphics[width=0.85\linewidth]{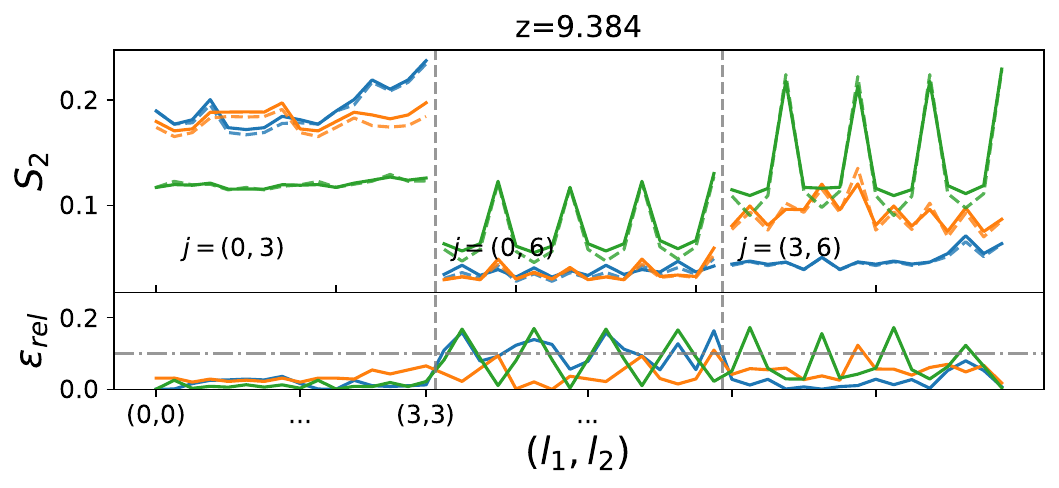}\\
\includegraphics[width=0.85\linewidth]{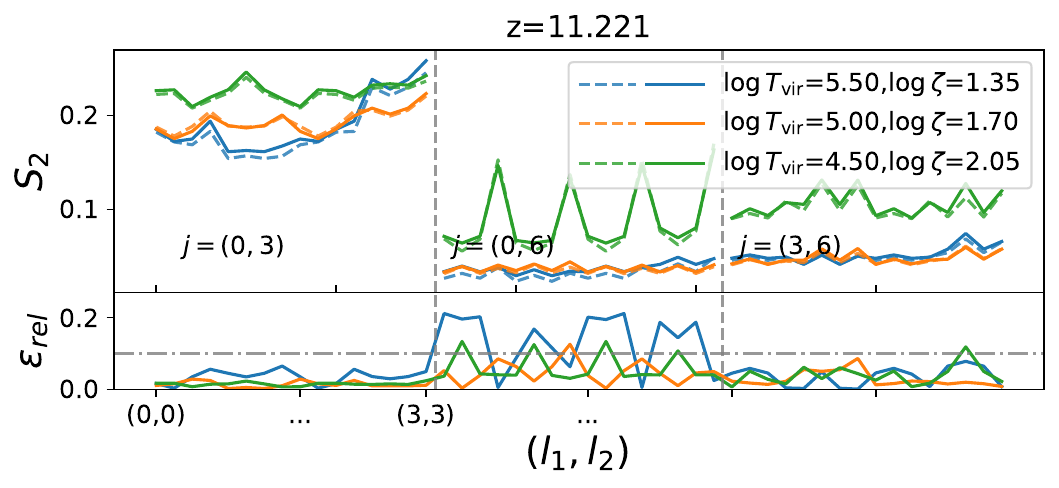}
\end{tabular}
\end{center}
\caption{Same as Figure~\ref{Fig:S264} but for the large-scale GAN and a choice of the index
$(j_1,j_2)$ taking the value of $(0,3)$, $(0,6)$ and $(3,6)$. Each set of reionization parameter is calculated with 1,024 clips of size $(2,256,128)$ from the raw image.}
\label{Fig:S2256}
\end{figure}

We show the comparison of ST coefficients of the large-scale GAN and the test set in Figure~\ref{Fig:S1256} (for $S_1$) and Figure~\ref{Fig:S2256} (for $S_2$), respectively. Here we set the $j=0,3,6$ to capture the large-scale information since the size of image sample is larger than in the case of small-scale GAN. 

For $S_1$, the relative error on small scales (i.e.\ small $j$) is small (about a few per cent), because our small-scale GAN has been well trained to provide reliable small-scale information. On the other hand, the relative error on large scales (i.e.\ large $j$) increases to $\sim 20\%$, a reasonable level of error given the very limited training set. For $S_2$ we find the similar trend. {An elevated error is observed for the $j=6, l=1$ coefficient in Figures~\ref{Fig:S1256} and \ref{Fig:S2256}. This coefficient corresponds to large-scale vertical features, indicating that artifacts caused by concatenation remains. The error excess in both statistics demonstrates the limitation of this method that the concatenating boundary can not be completely removed with limited training samples.}

\subsection{Test on Mode Collapse}

To assess the diversity of our large-scale GAN model, we implement several inspections, including visual inspection, pixel level variance, and feature level variance. 

\subsubsection{Visual inspection}
\begin{figure*}[h]
\centering
 \includegraphics[width=0.95\linewidth]{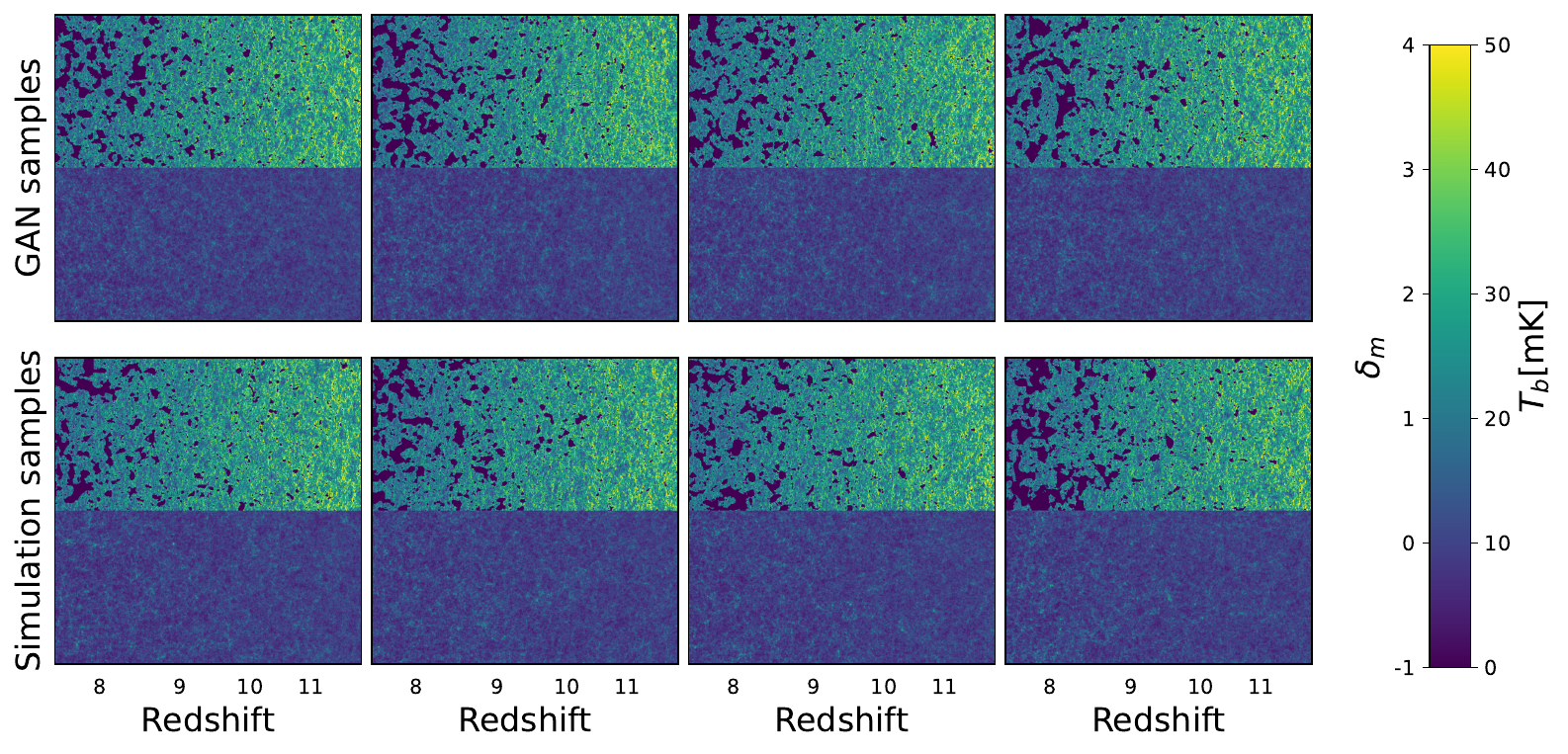}
 \caption{Visualization of the emulated images using the large-scale GAN (top), in comparison with the simulated images using 21cmFAST (bottom). Here we show four different realizations with the same parameters $(\log_{10} \zeta,\log_{10} T_{\rm vir}) = (1.7, 5.0)$. Each realization was computed using a different latent vector (for GAN) or initial condition (for simulation). In each panel, we show the 21~cm brightness temperature ($T_b$) field (the upper half) and the matter overdensity ($\delta_m$) field (the lower half). The LoS is along the x-axis.}
 \label{fig:vis}
\end{figure*}

We generate four realizations with the same set of reionization parameters for both GAN samples and simulation samples for visual inspection purposes, as illustrated in Figure~\ref{fig:vis}. We find that the shape and size of ionized bubbles exhibit variations across different GAN samples.
Furthermore, the locations of ionized bubbles also appear random, as no discernible trend or pattern is observed among the samples.

\subsubsection{Pixel level variance}
\begin{figure*}[h]
\centering
 \includegraphics[width=0.95\linewidth]{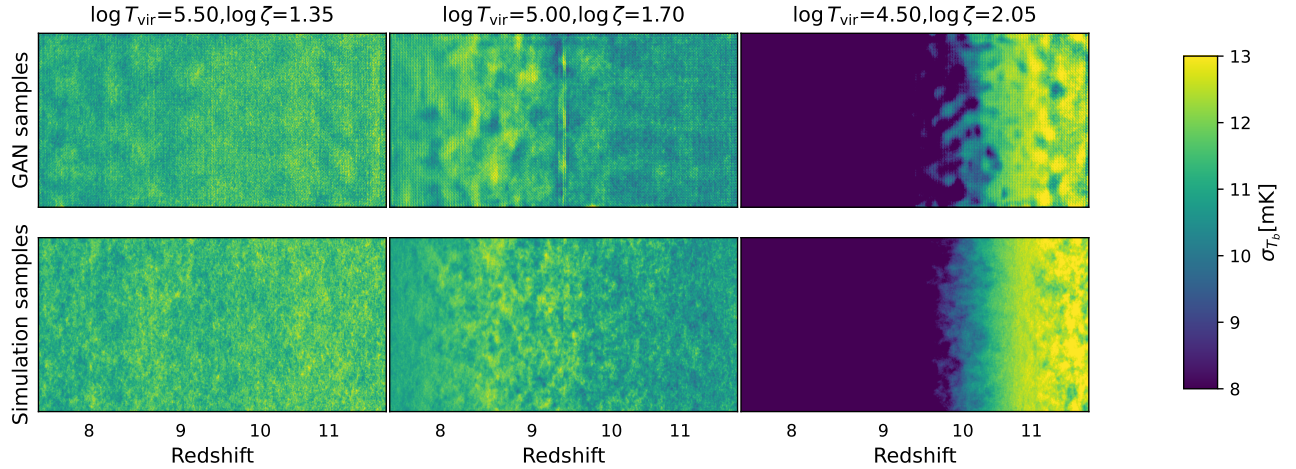}
 \caption{The standard deviation of the 21~cm brightness temperature map for each pixel over 1,024 image samples of the large-scale GAN (top), in comparison with the simulated images using 21cmFAST (bottom).}
 \label{fig:var}
\end{figure*}
We show the standard deviation of the $T_b$ field for each pixel over 1,024 image samples in Figure~\ref{fig:var}. Mode collapse would be indicated if the standard deviation in the large-scale GAN samples would be smaller than in the simulation test set. Figure~\ref{fig:var} shows that the variances for both GAN and test set samples appear similar, particularly when $T_b$ is large. Overall, we conclude that there is no evidence of significant mode collapse at the pixel level.


\subsubsection{Feature level variance}
\begin{figure*}
\begin{center}
\includegraphics[width=0.32\linewidth] {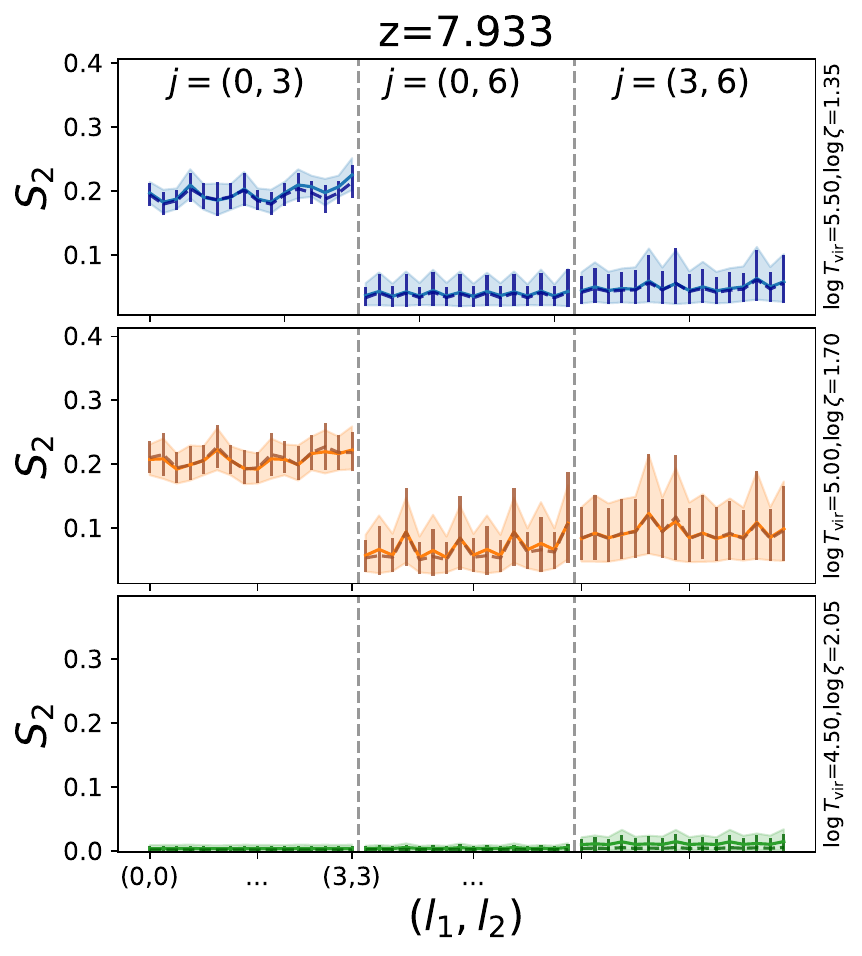}
\includegraphics[width=0.32\linewidth] {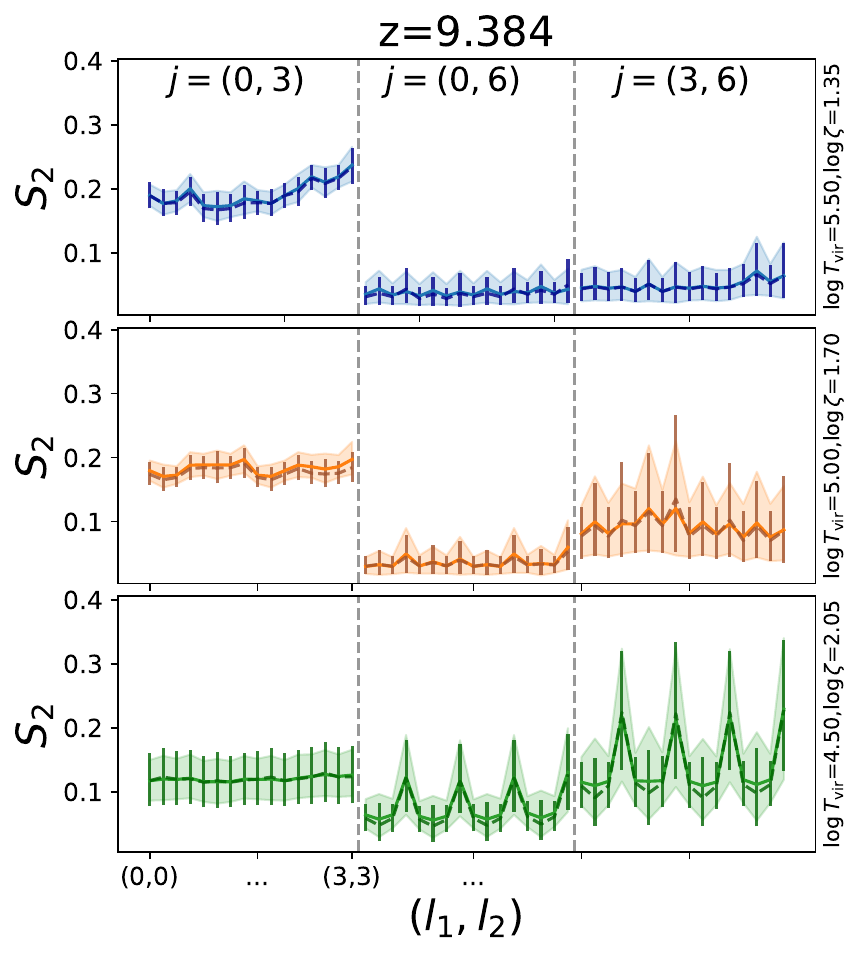}
\includegraphics[width=0.32\linewidth] {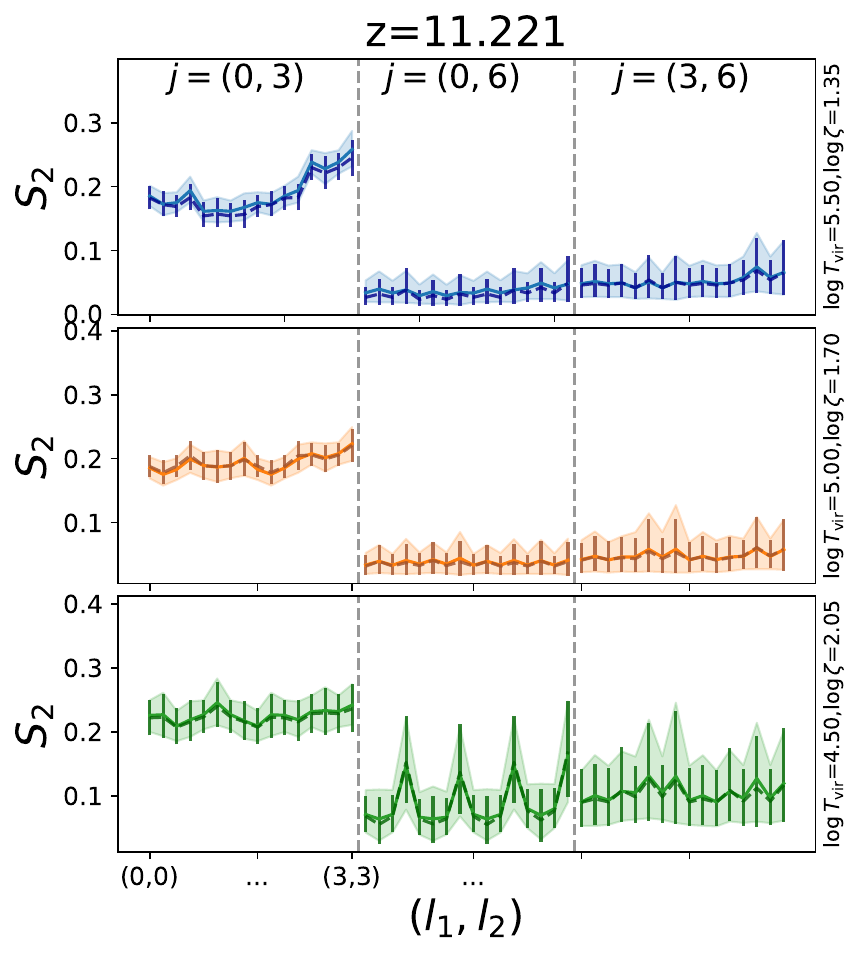}
\end{center}
\caption{The 2$\sigma$ scatter of the ST coefficient $S_{2}(j_1,l_1,j_2,l_2)$ of the 21~cm signal emulated with the large-scale GAN (dashed line for the mean and error bar for $2\sigma$ scatter) and that of the simulation test set images (solid line for the mean and shaded region for $2\sigma$ scatter) over 1,024 image samples at three representative redshifts of the center slice (from left to right) $z=7.933$, $9.384$, and $11.221$, respectively. We show the results with different values of reionization parameters (in different colors and rows). The indices $(j_1,j_2)$ take the value of $(0,3)$, $(0,6)$ and $(3,6)$.}
\label{Fig:S2std2561}
\end{figure*}
\begin{table*}
    \centering
    \caption{{Relative error of the mean value and 2$\sigma$ scatter for ST coefficients. Here the relative error is averaged over $j,l$ and three clips with the central redshift $z= \{7.933,9.384,11.221\}$. We choose $j={0,2,4}$ for the small-scale GAN, and $j=0,3,6$ for the large-scale GAN, in accordance with Figures \ref{Fig:S164}, \ref{Fig:S264}, \ref{Fig:S1256} and \ref{Fig:S2256}.}}
    \begin{tabular}{c|cccc|cccc}
           & \multicolumn{4}{c}{Small-scale GAN} & \multicolumn{4}{c}{Large-scale GAN} \\
$(\log T_{\rm vir},\log \zeta)$ &     $S_1$ mean  &    $S_1$ 2$\sigma $ &     $S_2$ mean  &    $S_2$ 2$\sigma $          &   $S_1$ mean  &    $S_1$ 2$\sigma $ &     $S_2$ mean  &    $S_2$ 2$\sigma $    \\
\hline
        $(5.5,1.35)$   &  1.5\%     &   1.7\%    &     1.6\%  &    1.8\%  &   4.7\%     &   5.6\%    &     5.2\%  &    4.7\%        \\
        $(5.0,1.70)$   &2.2\%     &   2.4\%    &     1.9\%  &    2.1\% &       3.2\%     &   4.0\%    &     3.6\%  &    3.8\%   \\ 
      $(4.5,2.05)$            &2.8\%     &   1.5\%    &     2.3\%  &    1.9\%   &       5.4\%     &   5.1\%    &     4.6\%  &    5.3\%   
\end{tabular}
    
    \label{tab:rel_err}
\end{table*}

We show in Figure~\ref{Fig:S2std2561} the 2$\sigma$ scatter (over 1,024 image samples) of the second-order ST coefficients $S_2$ of the $T_b$ field that serves as a representation of image feature. 
The $2\sigma$ scatter of GAN overlaps with that of simulation test set generically, indicating that there is no strong evidence of mode collapse at the feature level. The only exception is the case of 
$(\log_{10} \zeta,\log_{10} T_{\rm vir}) = (2.05, 4.5)$ at $z=9.384$ where disagreements in both the mean and $2\sigma$ scatter are found at large scales. 
This suggests a slight mode collapse issue in the generated images for that model at large scales at that particular redshift. 

{We also report the averaged relative error of both mean and $2\sigma$ scatter for $S_1$ and $S_2$ in Table \ref{tab:rel_err}. Our large-scale GAN exhibits two to three times larger error in both mean value and scatter than small-scale GAN. However, the error of $2\sigma$ keeps the same level as the mean value.} Overall, the GAN samples mimic the behavior of the simulation test set quite well, except for extreme cases (e.g.\ when $T_b$ is very small).

\subsection{Comparison of training set size to conventional training} \label{sec:fsd}

\begin{figure}
    \centering
    \includegraphics[width=\linewidth]{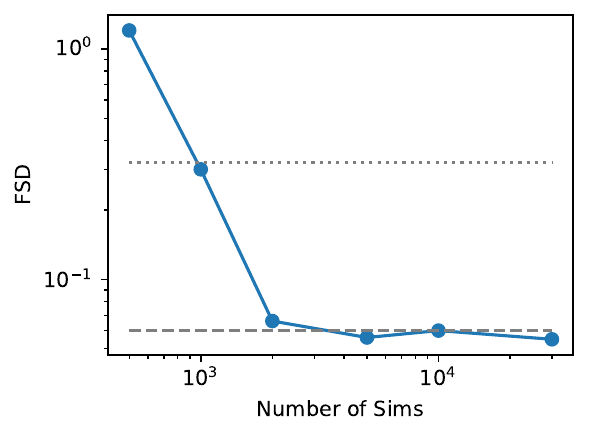}
    \caption{FSD for the small-scale GAN as a function of the number of training simulations (blue solid line). For comparison, the grey dashed line shows the FSD calculated using small-scale ST coefficients ($j=0,2,4$) for our trained large-scale GAN, while the grey dotted line shows the FSD calculated using large-scale ST coefficients ($j=2,4,6$) for the large-scale GAN.}
    \label{fig:fsd_baseline}
\end{figure}

{To assess the computational savings of our multi-fidelity approach compared to conventional GAN training, we first establish a baseline by evaluating the performance of a small-scale GAN trained with varying numbers of simulation samples. We quantify performance using the Fr\'echet Scattering Distance \citep[FSD;][]{2023MNRAS.526.1699Z}, which measures the similarity between two sets of samples (in this work, the GAN-generated and simulation data) based on the distance between the means and covariance matrices of their ST coefficients. The FSD is defined as}
\begin{equation} \label{eq:fsd}
\begin{aligned}
    {\rm FSD} = & \|\mu_{\rm GAN} - \mu_{\rm sim}\|^2 \\
               &+ {\rm tr}\left(\Sigma_{\rm GAN}+\Sigma_{\rm sim}-2(\Sigma_{\rm GAN}\Sigma_{\rm sim})^{1/2}\right)
\end{aligned}
\end{equation}
{where $\mu_{\rm GAN}$ and $\Sigma_{\rm GAN}$ are the mean vector and covariance matrix of ST coefficients derived from GAN samples, while $\mu_{\rm sim}$ and $\Sigma_{\rm sim}$ are those derived from the simulation samples. For this baseline analysis (small-scale GAN), we use ST coefficients with scales $j=0,2,4$, computed for three distinct patches along the redshift axis using test set reionization parameters, consistent with the coefficients presented in Figures~\ref{Fig:S164} and \ref{Fig:S264}. These coefficients are normalized by the mean values from the test set. We then average the FSD over different reionization parameters to get the final result.}

{The results for the small-scale GAN are shown in the solid blue line in Figure~\ref{fig:fsd_baseline}. The FSD decreases as the training set size increases but shows diminishing returns, plateauing after 5,000 simulations. This suggests that $\sim 5,000$ simulations are sufficient to train the small-scale GAN effectively.}

{We then evaluate our trained large-scale GAN (developed using the multi-fidelity approach). First, we compute its FSD using the same small-scale ST coefficients ($j=0,2,4$). The result (grey dashed line in Figure~\ref{fig:fsd_baseline}) achieves a low FSD comparable to the plateau value of the small-scale GAN, confirming that the large-scale GAN accurately reproduces the small-scale features. Meanwhile, we tested \{10,20,40,80\} largen-scale simulations as the high-fidelity set, confirmed that 80 simulations is the smallest number to keep a comparable small-scale FSD to the small-scale GAN after transfer learning. Secondly, we evaluate the large-scale GAN's performance on large scales using ST coefficients $j=2,4,6$. This `large-scale FSD', shown by the grey dotted line in Figure~\ref{fig:fsd_baseline}, is found to be $\sim 0.32$. Comparing this value to the baseline curve (blue solid line), the large-scale GAN's performance on large scales is at the same level achieved by the small-scale GAN when the latter is trained with $\sim 1,000$ simulations.}

{This comparison allows us to estimate the computational cost if we were to train the large-scale GAN conventionally (i.e., using only large-scale simulations). If we optimistically assume the required number of training samples does not scale with the output data size, it may still take $\sim 5,000$ large-scale simulations to reach a performance plateau, analogous to the small-scale case.}

{If we assume the FSD performance is proportional to the number of features, the number of large-scale features ($j=2,4,6$) in our large-scale simulations is $<1/4$ of the small-scale features ($j=0,2,4$). Therefore, reaching the same level of FSD requires four times more large-scale simulations.}
{With this estimated scaling, reaching an FSD level comparable to the small-scale plateau might necessitate $\gtrsim 4,000$ simulations just to match the current large-scale FSD performance ($\approx 0.32$). Reaching a fully converged, low FSD for large scales, potentially analogous to the small-scale plateau FSD, could possibly require $\gtrsim 20,000$ large-scale simulations if training conventionally. This highlights the substantial computational savings offered by our multi-fidelity training strategy.}

\section{Discussions and Conclusions}
\label{sec:con}

\begin{table*}
\centering
 \caption{Precision and Computational Cost for Various Methods}
 \label{tab:bench}
 \begin{tabular}{lccc}
 \hline\hline
  Method & \multicolumn{2}{c}{Relative Error}& Computational Cost\,\tablenotemark{\scriptsize{b}} \\
  \cline{2-3} 
   & At small scales & At large scales &  [$\times 10^4$ CPU core hours] \\
  \hline
  Small-scale GAN & $<10\%$ & --- & $1$\\
  Large-scale GAN (estimated\,\tablenotemark{\scriptsize{a}})&$<10\%$& $<10\%$&  $15 - 90$\\
 Large-scale GAN with few-shot transfer learning (this work) &$<10\%$ & $20\%-30\%$ &  $1.14$\\
  \hline
 \end{tabular}
 \flushleft
\tablenotetext{\scriptsize{a}}{Large-scale GAN is training GAN only with large-scale image samples from large-scale simulations, so its performance on precision and cost is estimated here.} 
\tablenotetext{\scriptsize{b}}{The computational cost here includes only the bottleneck cost in generating the data set, while the cost for training the GAN is cheap and not included in the table. For example, it only took us for 320 (10) GPU card hours to train the small-scale GAN (large-scale GAN with few-shot transfer learning) with four (two) GPU cards of Nvidia Tesla V100.}
\end{table*}

In this paper, we introduce the few-shot transfer learning technique as a realization of multi-fidelity emulation in ML. As an application, we build a GAN emulator for the large-scale 21~cm lightcone images. The multi-fidelity emulation involves a two-step process --- (1) building a StyleGAN2 emulator for small-scale images and training it with a huge number of training samples, and (2) modifying the model architecture to generate large-scale images and retraining the model with a limited number of training samples. 

Regarding computational cost, our multi-fidelity approach allows for building a large-scale GAN emulator with the cost of one to two orders of magnitude smaller than the naive GAN approach. Specifically, the training set in our paper comprises 120,000 image samples from 30,000 small-scale simulations and 320 image samples from 80 large-scale simulations, in a total of 11,400 CPU core hours for computational cost. If we were to build an emulator completely with training samples from large-scale simulations, it is estimated that at least 5,000 large-scale simulations would be required for training with about 150,000 CPU core hours. If for fair comparison purposes using the same amount of simulations to our paper, 30,000 large-scale simulations would cost 900,000 CPU core hours, about two orders of magnitude larger than our multi-fidelity approach. 

Regarding precision, our small-scale GAN emulates small-scale images with high precision, e.g.\ relative error generically less than $10\%$ for PS and $5\%$ for ST coefficients, and our large-scale GAN emulates large-scale images with reasonable precision, e.g.\ relative error $\gtrsim 30\%$ on very large scales $k\lesssim 0.02\,{\rm Mpc}^{-1}$ for PS and $\sim 20\%$ on large scales for ST coefficients, and on small scales with similar high precision to the small-scale GAN emulator. 

We summarize the precision and computational cost in Table~\ref{tab:bench}. In conclusion, our multi-fidelity approach can save $90\%-99\%$ computational cost in emulating high-quality images with reasonable precision. This implies that the few-shot transfer learning technique allows for emulating high-fidelity, traditionally computationally prohibitive, images in an economic manner. In principle, the application can be any two sets of highly correlated image training samples in low fidelity and high fidelity, respectively, e.g.\ small versus large scales (this work), low versus high resolutions, semi-numerical versus full-numerical simulations.  

{The application of multi-fidelity emulation approach will be particularly interesting for transfer learning from (low-fidelity) large-scale ($>500$ Mpc) semi-numerical simulations to (high-fidelity) small-scale ($\lesssim 100$ Mpc) fully-numerical hydrodynamic and radiative transfer simulations \citep[e.g.,][]{2022MNRAS.511.4005K,2014ApJ...793...29G}, to generate a large-scale emulator that contains both sophisticated astrophysical and hydrodynamic information on small scales and cosmological information on large scales. This super powerful emulator will be useful because large-scale fully-numerical simulations are computationally challenging. In this case, specific components might be adapted; for instance, the patchy-level discriminator could possibly be replaced by a large-scale discriminator incorporating a low-pass filter, which aims to stabilize the large-scale information while down-weighting small-scale details. While the required number of simulations will depend on the difference between high- and low-fidelity simulations, a similar number of simulations to this work could be sufficient.}

Note that some techniques herein can be further improved. Few-shot transfer learning may be realized by other techniques, e.g.\ based on mutual information or style vector CDC. Also, our multi-fidelity emulation may be applied to other generative models, e.g.\ normalizing flow, variational autoencoder, and diffusion model. We leave it to future work to explore other technical possibilities of multi-fidelity emulation. 

\section*{Acknowledgements}

This work is supported by the National SKA Program of China (grant No.~2020SKA0110401) and NSFC (grant No.~11821303). We thank Xiaosheng Zhao, Ce Sui, and Richard Grumitt for inspiring discussions. 
We acknowledge the Tsinghua Astrophysics High-Performance Computing platform at Tsinghua University for providing computational and data storage resources that have contributed to the research results reported within this paper. 

%

\vspace{5mm}


\software{\href{https://github.com/pytorch/pytorch}{PyTorch} \citep{Ansel_PyTorch_2_Faster_2024}, 21cmFAST \citep{2011MNRAS.411..955M,2020JOSS....5.2582M}, Kymatio \citep{Andreux_2020}, Matplotlib \citep{4160265}, numpy \citep{2020Natur.585..357H} }



\appendix

\section{Details of GAN architecture and training configurations}
\label{app:GAN}

\begin{figure}
\begin{center}
\includegraphics[width=0.85\linewidth]{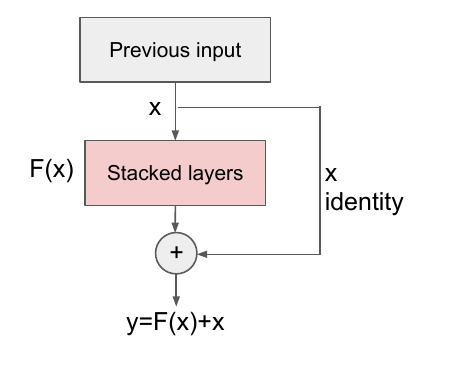}
\end{center}
\caption{An illustration of a ResNet block.}
\label{Fig:resnet}
\end{figure}

In this appendix, we present the network structure in detail. 

\subsection{Generator}

For our small-scale GAN, as described in Section \ref{sec:GAN}, the generator consists of a mapping network $f$ and a synthesis network $g$. The mapping network $f$ is constructed by two multi-layer perceptions (MLP). One MLP is an eight-layer one $f_1$, mapping a Gaussian random vector $\mathbf{z}$ to a vector of length 512: $f_1(\mathbf{z})$. The other MLP is a two-layer MLP $f_2$, mapping astrophysical parameter $\mathbf{c}$ to a 256-length vector $f_2(\mathbf{c})$. Then half of the components in the $f_1(\mathbf{z})$ are multiplied by $f_2(\mathbf{c})$ to form the final style vector $\mathbf{w}$. For the synthesis network $g$, it starts from a fixed layer of the size $(512,2,16)$, then convolved twice before a two-times upsampling. Right after each convolution, Gaussian noise of the same size is injected into the feature map. After five times of upsampling, the feature map increases from size $(512,2,16)$ to $(256,64,512)$, and the reduction in channels is to save memory usage. Before each upsampling, an additional convolution layer converts the current feature map to a final image with the corresponding size, e.g. before the first convolution, the layer converts the feature map of size $(512,2,16)$ to a pre-final image of size $(2,2,16)$. By upsampling all pre-final images to the final size and adding them together, we obtain the final output image of size $(2,64,512)$.

The style vector $\mathbf{w}$ shapes the convolutional weights as follows. The convolution kernel can be expressed by a 4-dimensional tensor $k_{ijkl}$, where $i$ is the input channel, $j$ is the output channel, $k$ and $l$ are spatial indices. The tensor $k_{ijkl}$ is normalized with the style vector $\mathbf{w}$, 
\begin{equation}
    \begin{aligned}
        k_{ijkl}^\prime &= \mathbf{w}_i \cdot k_{ijkl}\,,\\
        k_{ijkl}^{\prime\prime} &= k_{ijkl}^\prime \bigg{/}\sqrt{\sum_{i,k,l}k_{ijkl}^{\prime}{}^2+\epsilon}\,,
    \end{aligned}
\end{equation}
where $\epsilon$ is a small number to avoid numerical error. 

\subsection{Discriminator}

The discriminator is constructed using an input layer, five ResNet blocks, and a two-layer MLP. We illustrate a ResNet block in Figure~\ref{Fig:resnet}. The input data is added to the result of the stacked layers that is the input of the next layer. In our settings, each block has two convolution layers as the stacked layer in this block. Between each ResNet block, we downsample the feature map by a factor of two. For the small-scale GAN, after five ResNet blocks, the spatial dimension of the feature map drops from $(64,512)$ to $(4,32)$. Then the map is flattened to be a long vector. The MLP has two inputs --- the long vector and astrophysical parameters, and its output is a score for which zero means real and unity means fake.

\subsection{Training configurations}
\label{app:trainconf}

For the small-scale GAN, we employ four GPU cards of Nvidia Tesla V100. The training {is carried out with} a batch size of 32 in each card, in a total of approximately 320 GPU card hours. 

For the large-scale GAN, we employ two GPU cards of Nvidia Tesla V100. We run the training for 4,000 iterations with a batch size of four in each card. The cost for training the large-scale GAN is approximately 10 GPU card hours.

The hyperparameters we choose can be found in our GitHub repo.

\subsection{Convergence}
\label{app:cov}
\begin{figure}
    \centering
    \includegraphics[width=\linewidth]{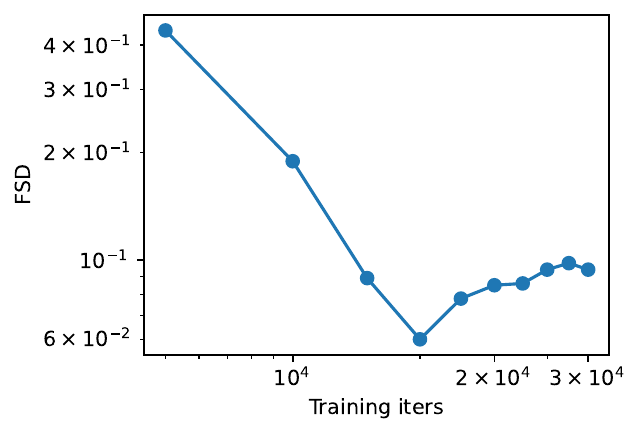}
    \caption{FSD with $j=0,2,4$ for the small-scale GAN with 10,000 simulations as a function of the training iterations.}
    \label{fig:fsd_training}
\end{figure}
{In traditional machine learning models, convergence is often achieved when a loss function reaches a minimum. However, for GAN,  convergence is significantly more complex and harder to define definitively due to the adversarial nature and non-single objective. Practically, the GAN convergence is ensured by monitoring quantitative metrics and picking the training iteration with the best metric \citep[e.g. Figure 1 in][]{2020arXiv200606676K}. Similar to the Frechét Inception score which is commonly used in computer vision, we monitor the FSD during the training. An example of the evolution of FSD with 10,000 training simulations is shown in Figure \ref{fig:fsd_training}, which shows a clear minimum at around 20,000 iterations.}

{For the small-scale GAN, we monitor the FSD with $j=0,2,4$ after every 5,000 iterations, and found that at 40,000 iterations it reaches best performance. For the large-scale GAN, we adopt the FSD with $j=2,4,6$ to capture the performance on large scales and monitor the FSD after every 200 iterations. We found that at 1,400 iterations it gives the optimal performance with 80 large-scale training simulations.}

\section{Attempted Conventional Training with 80 large-scale simulations}
\label{app:80}
\begin{figure*}
    \centering
    \includegraphics[width=0.9\linewidth]{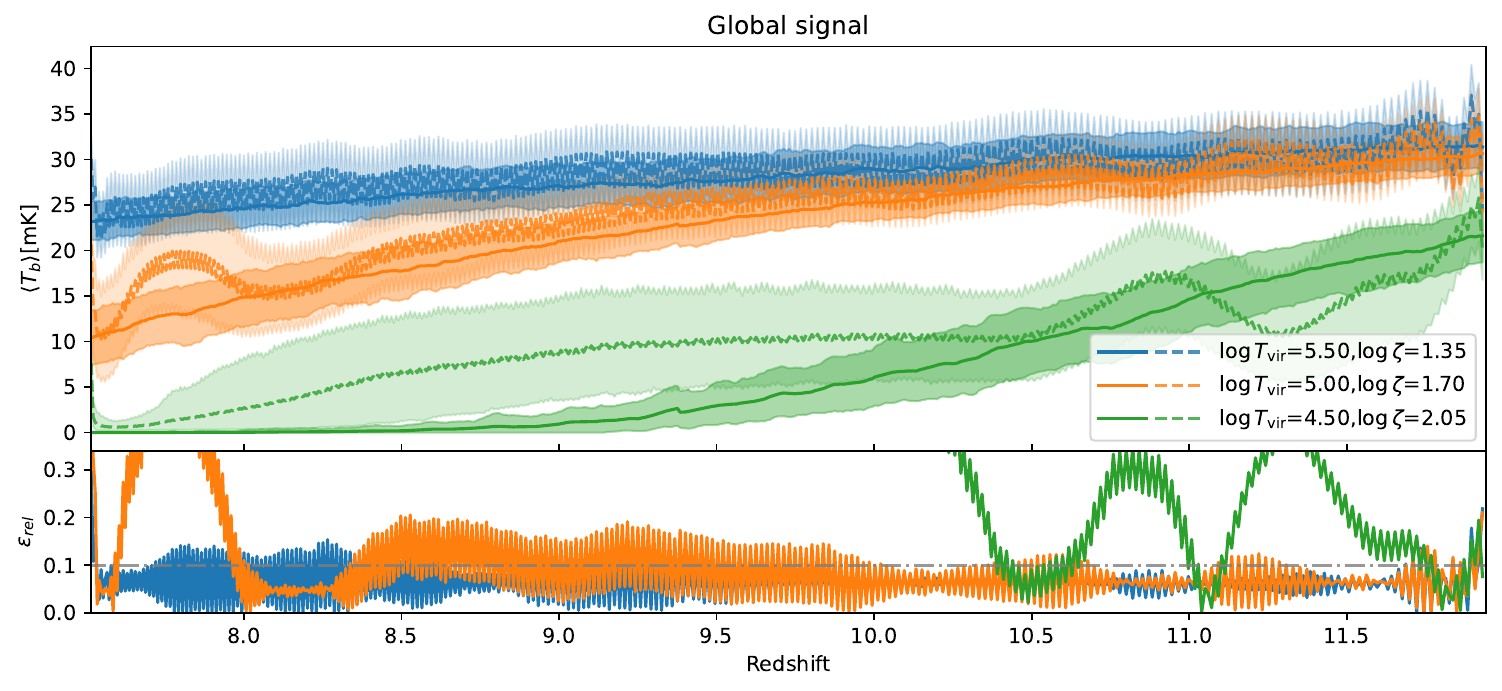}
    \caption{{Same as Figure~\ref{fig:glob64} but for the large-scale GAN trained conventionally using only 80 simulation samples.}}
    \label{fig:glob80}
\end{figure*}

\begin{figure*}
    \centering
    \includegraphics[width=0.95\linewidth]{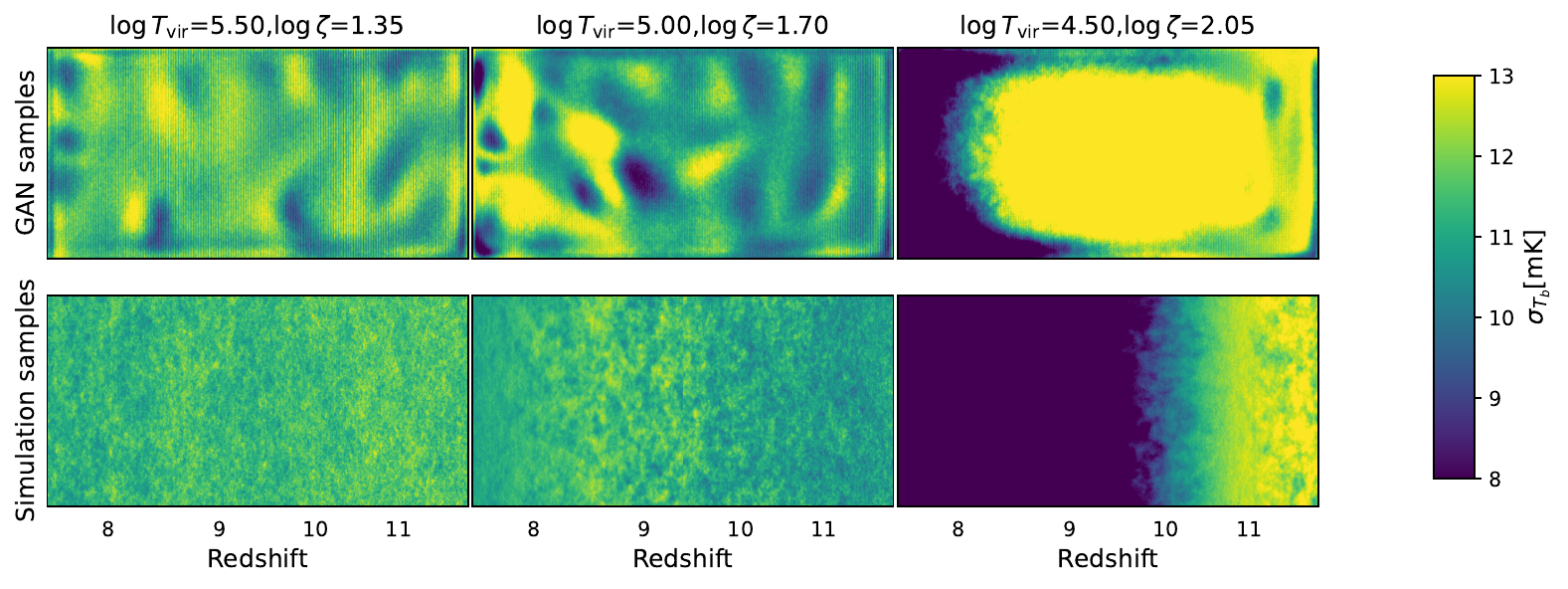}
    \caption{{Same as Figure~\ref{fig:var} but for the large-scale GAN trained conventionally using only 80 simulation samples.}}
    \label{fig:var_80}
\end{figure*}

{We attempt to train the large-scale GAN conventionally, using a limited dataset of only 80 large-scale simulations. The results demonstrate the inadequacy of this approach with insufficient training data. Figure~\ref{fig:glob80} shows the predicted global 21-cm signal, revealing significant discrepancies compared to the true signal, particularly for the early reionization model (e.g., log $T_{\rm vir}=5.50$, log $\zeta=1.35$). Furthermore, the presence of unphysical oscillations (``wiggles'') in the predicted global signal suggests potential mode collapse. This is strongly corroborated by examining the pixel-level variance shown in Figure~\ref{fig:var_80}. The GAN fails to reproduce the variance trends seen in the simulations, exhibiting erratic behavior indicative of mode collapse. Due to this poor performance and clear evidence of training instability, we exclude the results using this conventionally trained, data-limited GAN from the comparisons in the main body of this paper.}


\bibliography{reference}{}
\bibliographystyle{aasjournal}



\end{document}